\pdfoutput=1
\documentclass[aps,prl,floatfix,superscriptaddress,reprint]{revtex4-2}
\usepackage{eurosym}
\usepackage{amsbsy}
\usepackage{latexsym,epsfig,graphicx}
\usepackage{dcolumn}
\usepackage{subfigure}
\usepackage{comment}
\usepackage{color}
\usepackage{bm}
\usepackage{mathrsfs}
\usepackage{amssymb}
\usepackage{amsfonts}
\usepackage{amsmath}
\usepackage{xspace}
\usepackage{epstopdf}
\usepackage{tabularx}
\usepackage{longtable}
\usepackage[colorlinks=true, letterpaper=true, pdfstartview=FitV, linkcolor=blue, citecolor=blue, urlcolor=blue]{hyperref}
\usepackage[normalem]{ulem}
\usepackage[version=4]{mhchem}

\begin{document}
\title{Exceptional Non-Abelian Topology in Multiband Non-Hermitian Systems}

\author{Cui-Xian Guo}
\affiliation{Beijing National Laboratory for Condensed Matter Physics, Institute of Physics, Chinese Academy of Sciences, Beijing 100190, China}
\author{Shu Chen}
\affiliation{Beijing National Laboratory for Condensed Matter Physics, Institute of Physics, Chinese Academy of Sciences, Beijing 100190, China}
\affiliation{School of Physical Sciences, University of Chinese Academy of Sciences, Beijing 100049, China}
\affiliation{Yangtze River Delta Physics Research Center, Liyang, Jiangsu 213300, China}
\author{Kun Ding}
\affiliation{Department of Physics, State Key Laboratory of Surface Physics, and Key Laboratory of Micro and Nano Photonic Structures (Ministry of Education), Fudan University, Shanghai 200438, China}
\author{Haiping Hu}
\thanks{Corresponding author: hhu@iphy.ac.cn}
\affiliation{Beijing National Laboratory for Condensed Matter Physics, Institute of Physics, Chinese Academy of Sciences, Beijing 100190, China}
\affiliation{School of Physical Sciences, University of Chinese Academy of Sciences, Beijing 100049, China}
\begin{abstract}
Defective spectral degeneracy, known as exceptional point (EP), lies at the heart of various intriguing phenomena in optics, acoustics, and other nonconservative systems. Despite extensive studies in the past two decades, the \textit{collective} behaviors (e.g., annihilation, coalescence, braiding, etc.) involving multiple exceptional points or lines and their interplay have been rarely understood. Here we put forward a universal non-Abelian conservation rule governing these collective behaviors in generic multiband non-Hermitian systems and uncover several counterintuitive phenomena. We demonstrate that two EPs with opposite charges (even the pairwise created) do not necessarily annihilate, depending on how they approach each other. Furthermore, we unveil that the conservation rule imposes strict constraints on the permissible exceptional-line configurations. It excludes structures like Hopf link yet permits novel staggered rings composed of noncommutative exceptional lines. These intriguing phenomena are illustrated by concrete models which could be readily implemented in platforms like coupled acoustic cavities, optical waveguides, and ring resonators. Our findings lay the cornerstone for a comprehensive understanding of the exceptional non-Abelian topology and shed light on the versatile manipulations and applications based on exceptional degeneracies in nonconservative systems.
\end{abstract}
\maketitle

Exceptional points (EPs) are peculiar spectral singularities induced by non-Hermiticity \cite{epberry,ep,ep2,nhbook}. The past decade has witnessed a myriad of remarkable phenomena and functionalities in optics, photonics, and acoustics pivoted on the non-Hermitian degeneracy \cite{nhreview,nhreview2}, such as single-mode lasing \cite{lasing1,lasing2}, unidirectional transmission or reflection \cite{ptreflection1,ptreflection2,wangx}, enhanced sensing \cite{sense2,sense3,sense4,sense5}, and unconventional quantum interference or correlation \cite{Longhir,Caor,Roccatir,Gongr}. Unlike Dirac or Weyl point in Hermitian systems, both the eigenenergies and eigenvectors coalesce at an EP. Without symmetry constraints, an EP of second order is stable in two-dimensional (2D) parameter space \cite{Yang} and extends to exceptional line (EL) in 3D \cite{epline1,epline2,epline3,epline4,epline5,epline7,epline8,epline10,epline11,jphknot,werfansh}.

From a ``local" perspective, the simplest EP is dictated by a nondiagonalizable two-by-two Hamiltonian whose eigenvalues have a square-root singularity. The EP can be assigned a topological charge \cite{epcharge,epscience,Fuliang}, or discriminant number \cite{Yang} that signifies the eigenvalue permutation \cite{ep2encircling,ep2encircling2} and ensures its stability against small perturbations. While most of the aforementioned phenomena are well understood by scrutinizing one single EP, in most generic non-Hermitian settings, multiple exceptional degeneracies may emerge, annihilate, coalesce, and braid with varying system parameters. Thus far, a holistic framework governing these collective behaviors involving multiple EPs (or ELs) and their interplay in multiband systems remains elusive. What new interesting physics is nurtured by multiple EPs or ELs beyond their local descriptions? And are there any ``emergent" phenomena intrinsic to multiband systems beyond the two-band case? Addressing these questions not only provides a fundamental understanding of non-Hermitian physics, but also sheds light on the manipulations and functional design of exceptional degeneracies relevant in a wide range of nonconservative systems like coupled ring resonators \cite{fansh1,fansh2,ringcavity1,ringcavity2,ringcavity3,ringcavity4}, optical waveguides \cite{waveguide1,waveguide2,waveguide3,encircleEP5,encircleEP7}, acoustic cavities \cite{ep2encircling3,epnexus,pg3,hep1,hep2}, or photonic quantum walks \cite{xue1,xue2}.

In this Letter, we demonstrate that the collective behaviors and parametric evolution of multiple EPs or ELs are governed by a universal non-Abelian conservation rule (NACR). From the rule, we uncover two intriguing and counterintuitive phenomena. Firstly, it is usually taken for granted that two EPs with opposite charges annihilate each other. In stark contrast, we show that an EP pair (even the pairwise created) in multiband non-Hermitian systems does not necessarily annihilate. Their annihilation or coalescence is path dependent and exhibits an ``adjacent" effect. Secondly, as stereotyped in the two-band case, either the nodal lines \cite{Zhong_2017,Yan_Hopf,Bi_knot,Ezawa_Hopf,Chen_Hopf,XGWan,jphknot,Lee_knot,epline11} or ELs \cite{epline1,epline2,epline3,epline4,epline5,epline7,epline8,epline10,epline11,jphknot,werfansh} can form any desired configurations (e.g., Hopf link, trefoil knot, etc.). In multiband settings, the NACR puts strict constraints on the permissible configurations and evolution of ELs. For instance, structures like Hopf link composed of noncommutative ELs are forbidden, while novel staggered exceptional rings are allowed. We further propose a three-state system readily realizable in various experimental platforms (e.g., acoustic cavities) to observe these phenomena. We emphasize that these unexpected results are a consequence of the underlying non-Abelian topology and intrinsic to multiband non-Hermitian systems.

\begin{figure}[!t]
\includegraphics[width=3.35in]{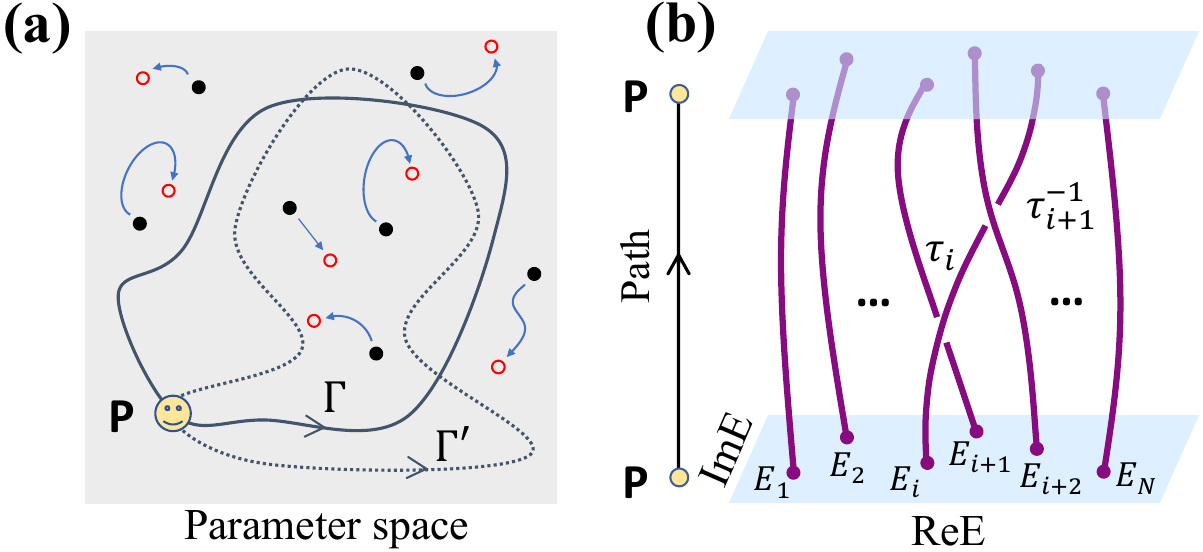}
\caption{Schematics of the NACR and braid invariant. (a) Sketch of closed paths based at $\bm{P}$ (start and end point) enclosing multiple EPs (black dots) in the 2D parameter space. The (black) arrow marks the direction of the path. The analysis equally applies to ELs in the 3D parameter space. Path $\Gamma$ (solid line) and $\Gamma'$ (dotted line) are topologically equivalent by smooth deformation. With varying system parameters, the EPs are shifted from their initial positions to final positions (red circles), as marked by blue arrows. (b) An exemplary braiding of eigenenergy strands of an $N$-band non-Hermitian system along a closed path based at $\bm{P}$.}
\label{fig1}
\end{figure}
{\color{blue}\textit{Non-Abelian conservation rule (NACR).—}}We consider a generic $N$-band non-Hermitian system. The parameter space is punctured by some exceptional degeneracies [EPs (ELs) in 2D (3D)], as sketched in Fig. \ref{fig1}(a). To address their collective behaviors, we investigate the closed paths with a base point $\bm{P}$ (start and end point) enclosing these degeneracies in the parameter space. Such closed paths are characterized by the fundamental group of the Hamiltonian space $X_N$ \cite{braidclass1,braidclass2,Hu1,Hu2,FSH,Bzduek,SM}:
\begin{eqnarray}
\pi_1(\bm{P},X_N)=B_N\label{pi1}.
\end{eqnarray}
$B_N$ is the braid group. Thus each path is assigned a braid-valued topological invariant. It describes how the complex eigenenergies evolve along the path. $B_N$ is non-Abelian except for $N=2$ with $B_2=\mathbb{Z}$, wherein the braid invariant is the discriminant number \cite{Yang}. Figure \ref{fig1}(b) depicts a representative eigenlevel braiding along some closed path. A convenient way to obtain the braid invariant is through Artin's word. After sorting the real parts of eigenenergies as $\textrm{Re}E_1\leq \textrm{Re}E_2\leq...\leq \textrm{Re}E_N$, the $i$th level crosses over or under the $(i+1)$th level is marked as $\tau_i$ or $\tau_i^{-1}$. Any braid-group element is represented as a sequence of over and under crossings [e.g., $\tau_i\tau_{i+1}^{-1}$ in Fig. \ref{fig1}(b)]. $\tau_i$'s satisfy the braid relations
\begin{eqnarray}\label{braidrelation}
\Big\{\begin{array}{l l} \tau_i\tau_j=\tau_j\tau_i,~~~~~~~~~~~~~~~~~\quad \textrm{if}~|j-i|>1; \\ \tau_i\tau_{i+1}\tau_i=\tau_{i+1}\tau_i\tau_{i+1}, \quad \textrm{any}~1\leq i\leq N-1.\\ \end{array}
\end{eqnarray}

The homotopy theory \cite{Mermin,Teo} immediately implies that a smoothly morphing path without touching any EPs or ELs, e.g., $\Gamma\rightarrow\Gamma'$ as in Fig. \ref{fig1}(a), yields the same braid invariant. It can be regarded as the NACR for static non-Hermitian Hamiltonians. The flow conservation \cite{elchain}, non-Hermitian doubling theorem \cite{Yang}, and no-go theorem \cite{Hu2} are the special cases of this static NACR \cite{SM}. We proceed to consider a time-varying Hamiltonian $H[\bm{\lambda}(t)]$ with parameter $\bm{\lambda}$. We investigate the stroboscopic evolution \cite{ep2encircling,ep2encircling2,ep2encircling3,stroencircle1,stroencircle2,pg1} of EPs or ELs wherein the nonadiabatic transitions typically encountered in dynamic evolutions can be avoided, and focus on a fixed path $\Gamma$ in the parameter space, as sketched in Fig. \ref{fig1}(a). It can be shown that as long as no EPs or ELs pass through the path $\Gamma$ during the whole evolution, the braid invariants at the initial time $b_{\Gamma}(t_i)$ and final time $b_{\Gamma}(t_f)$ are conjugate,
\begin{eqnarray}\label{dcr}
b_{\Gamma}(t_f)=b_{dyn}^{-1}~b_{\Gamma}(t_i)~b_{dyn}.
\end{eqnarray}
Here $b_{dyn}$ is purely a dynamical factor describing the accumulated braiding of (instantaneous) eigenenergy from time $t_i$ to $t_f$ at the base point $\bm{P}$ \cite{SM}. As the factor $b_{dyn}$ acts indiscriminately on all the closed paths based at $\bm{P}$, it would not affect the non-Abelian properties of multiple EPs or ELs. By suitably choosing the base point, we can set $b_{dyn}=1$. We dub Eq. (\ref{dcr}) as a dynamical NACR under parametric evolution. The braid invariant may change during the evolution once extra EPs (ELs) enter or leave the path. As will be seen later, this ostensibly simple rule is powerful in analyzing the collective phenomena of multiple EPs or ELs.

\begin{figure}[btp]
\includegraphics[width=3.35in]{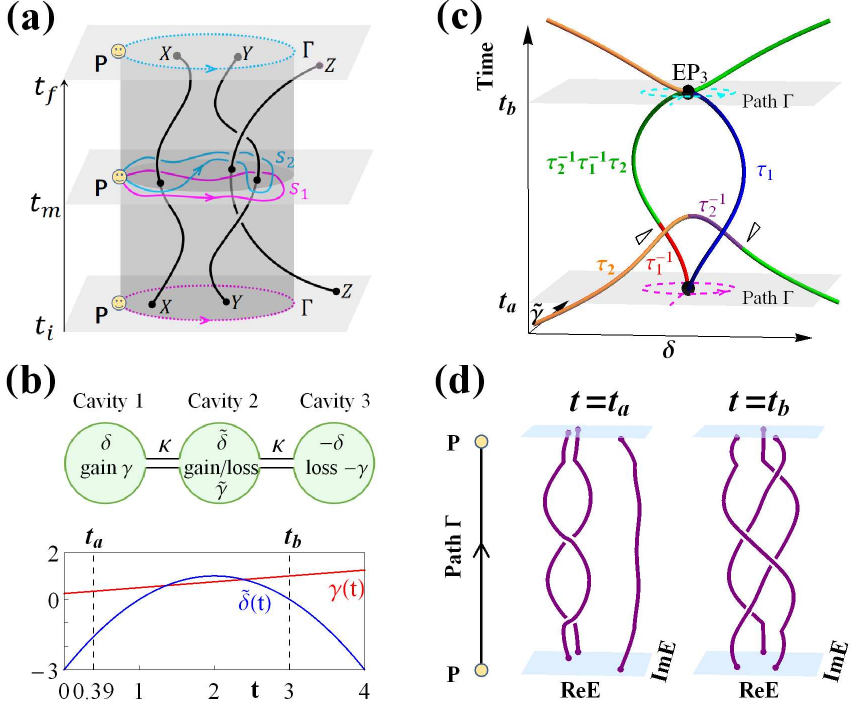}
\caption{Annihilation and coalescence of EPs in multiband non-Hermitian systems. (a) Time-evolution loci of the three EP $X$, $Y$, $Z$ in the parameter space (black curves). Path $\Gamma$ at the initial (final) stage is topologically equivalent to $s_1$ ($s_2$). (b) (top) Schematics of three coupled acoustic cavities to realize the model (\ref{H2D}). $\kappa$ is the coupling strength. $\delta$, $\tilde{\delta}$ are the detunings. (bottom) Control parameters $\gamma(t)$ and $\tilde{\delta}(t)$ of the protocol as a function of time $t$. Here $\gamma(t)=(t+1)/4$, $\tilde{\delta}(t)=1-(t-2)^2$. (c) EP loci of the model (\ref{H2D}) in the 2D $(\delta, \tilde{\gamma})$ space. The EPs are marked by different colors according to their braidings. $t_a\approx0.39$ and $t_b=3$ are the time instants when the EP pair emerges and coalesces. The sudden changes of colors are marked by black triangles. (d) The eigenvalue braidings along the path $\Gamma$ in (c) at $t=t_a$ and $t=t_b$. For (c) and (d) the base point $\bm{P}$ is pinned at $(\delta,\tilde{\gamma})=(0,-3)$.}
\label{fig2}
\end{figure}

{\color{blue}\textit{Annihilation and coalescence of EPs.—}}The non-Abelian exceptional topology brings key nonlocal features in the merging, annihilation, and coalescence process of EPs. As the first application of the NACR, we investigate the merging of two EPs with opposite charges. Figure \ref{fig2}(a) sketches a bizarre case of two EPs (labeled as $X$ and $Y$) in the parameter space with a time-varying Hamiltonian. $Y$ bypasses another EP (labeled as $Z$) before rejoining with $X$. For this case, $Z$ enters and then leaves the closed path $\Gamma$ during the process. Suppose $X$ and $Y$ were initially created pairwise from a Dirac point or a hybrid EP  \cite{Fuliang,dptoep1,dptoep2,hep1,hep2,hep3,SM}. The local braidings of $X$, $Y$, $Z$ are denoted as $b_X$, $b_Y$, and $b_Z$, respectively. We have the initial braiding $b_{\Gamma}(t_i)=b_Xb_Y=1$ and final braiding $b_{\Gamma}(t_f)=b_Xb_Z^{-1}b_X^{-1}b_Z$ \cite{SM}. The two EPs do not annihilate each other eventually, except when $b_X$ and $b_Z$ commute, $b_Xb_Z=b_Zb_X$. Otherwise, they would coalesce into a higher-order EP. To visualize the difference, we note that the path $\Gamma$ at $t_i$ and $t_f$ is smoothly deformed to topologically distinct paths $s_1$ and $s_2$ at some intermediate time [Fig. \ref{fig2}(a)]. As the noncommutativity occurs between neighboring braidings from the relation in Eq. (\ref{braidrelation}), the annihilation and coalescence exhibit an ``adjacent" effect, where an EP pair between the $i$th and $(i+1$)th bands (with braiding $\tau_i^{\pm1}$) is unaffected (affected) by its nonadjacent (adjacent) EP (with braiding $\tau_{i\pm1}^{\pm}$).

{\color{blue}\textit{Experimental realization.—}}The path-dependent annihilation of EPs is best illustrated by the following three-state model:
\begin{equation}\label{H2D}
H=\left(
\begin{array}{ccc}
\sqrt{2}i[\gamma(t)+i\delta] & -\kappa & 0 \\
-\kappa & i[\tilde{\gamma}+i\tilde{\delta}(t)] & -\kappa \\
0 & -\kappa & -\sqrt{2}i[\gamma(t)+i\delta] \\
\end{array}
\right).
\end{equation}
The model can be readily implemented in various experimental platforms, e.g., coupled acoustic cavities \cite{ep2encircling3,epnexus,pg3,hep1,hep2}, as depicted in Fig. \ref{fig2}(b). Here $\kappa$ is the coupling strength between neighboring cavities. $\kappa=1$ is set as the energy unit. $\delta$, $\tilde{\delta}(t)$, $-\delta$ are the detunings and $\gamma(t)$, $\tilde{\gamma}$, $-\gamma(t)$ are the gain or loss in the respective cavities. We note that the main physics stays unchanged if only the loss term is present  \cite{Ornigotti1r, Joglekarr}. 

We vary the system parameters as $\gamma(t)=(t+1)/4$ and $\tilde{\delta}(t)=1-(t-2)^2$ and examine the evolution of EPs in the 2D $(\delta,\tilde{\gamma})$ space \cite{SM}. Figure \ref{fig2}(c) plots the EP loci, with different colors marking their braid-valued invariants \cite{SM}. In acoustic cavities, the EP loci can be extracted by measuring the pressure response spectra. Targeted on a pair of EPs created at $t_a\approx0.39$ with opposite braidings $\tau_1$ and $\tau_1^{-1}$, we observe their subsequent detouring, merging at $t_b=3$, and splitting for $t>t_b$. Note the abrupt change of braid invariant [marked by the black triangle in Fig. \ref{fig2}(c)] from $\tau_1^{-1}$ (red) to $\tau_2^{-1}\tau_1^{-1}\tau_2$ (green) when the EP undercrosses another EP with braiding $\tau_2$ (orange). This is due to the noncommutativity between the braidings $\tau_1^{-1}$ and $\tau_2$ \cite{SM}. Instead of annihilation, the two initial EPs merge into a third-order EP at $t_b=3$. Figure \ref{fig2}(d) shows the eigenenergy braidings associated with the path $\Gamma$ at the two time instants $t=t_a$ (when they are created) and $t=t_b$ (when they coalesce). The braid invariant is $1$ (trivial) for the former and $\tau_1\tau_2^{-1}\tau_1^{-1}\tau_2$ for the latter, in agreement with their nonannihilation at $t=t_b$. In experiments, the different properties of the two merging points can be extracted by measuring the eigenspectra nearby or the phase rigidity \cite{epnexus}.

\begin{figure}[btp]
\includegraphics[width=3.35in]{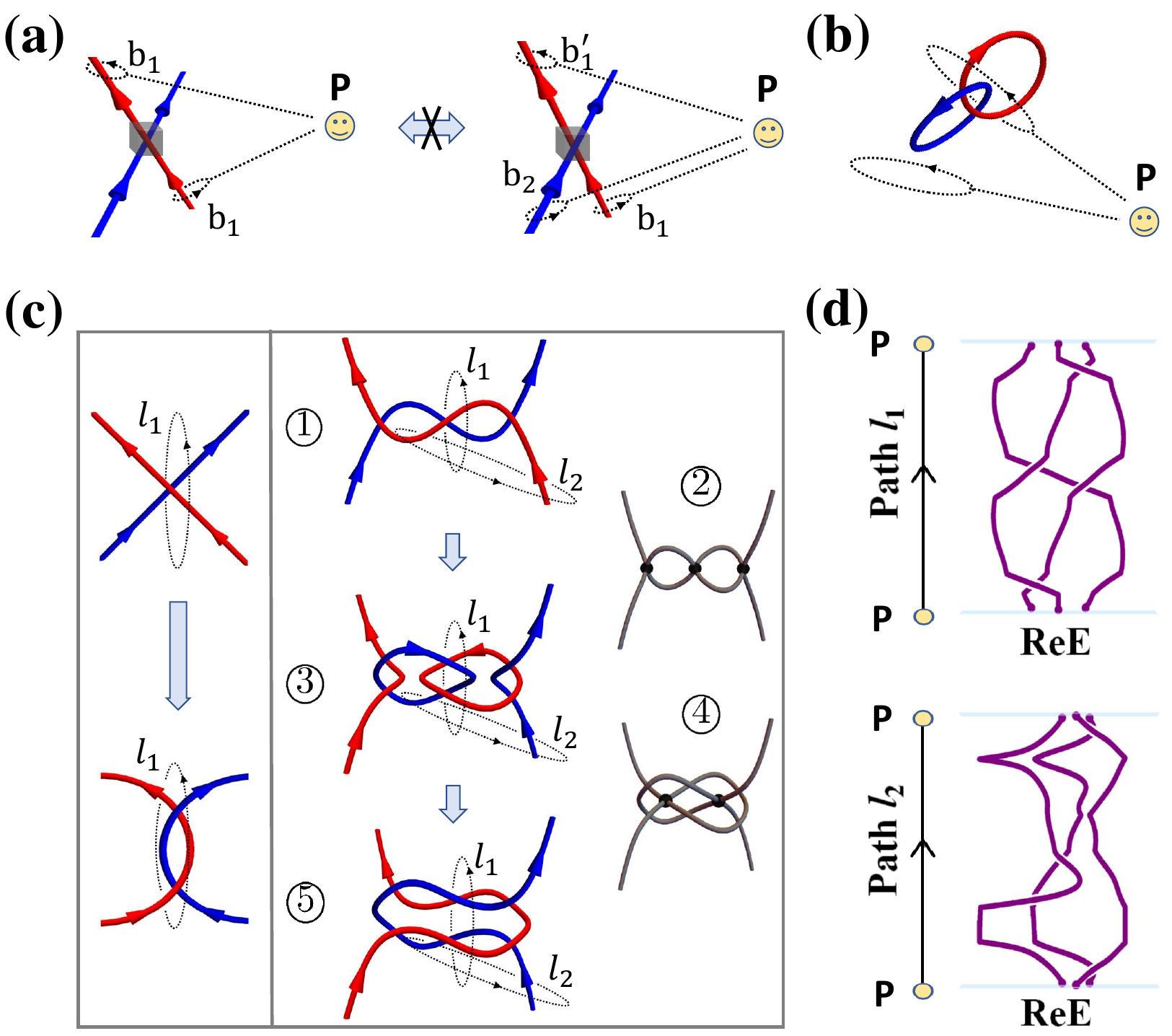}
\caption{Admissible exceptional lines (ELs) constrained by the NACR. (a) Two different configurations of ELs (red and blue lines). The dotted lines denote the encircling paths based at $\textrm{P}$ with their braid invariants labeled. (b) Hopf link of non-commutative ELs as a no-go structure. (c) The permissible evolution process of two non-commutative ELs (in red and blue) for model (\ref{H3D}). The touchings in step \textcircled{2} are through higher-order EPs (black dots). The braid invariants stay unchanged in all the steps for path $l_1$ and $l_2$ as per the NACR. (d) the eigenvalue braidings for path $l_1$ and $l_2$ at step \textcircled{3}. The arrows of the ELs in (a)(b)(c) mark the flow.}
\label{fig3}
\end{figure}

{\color{blue}\textit{Admissible ELs by the conservation rule.—}}In 3D, the exceptional non-Abelian topology manifests as permissible EL structures compatible with the NACR. To gain intuition, Fig. \ref{fig3}(a) shows two configurations with the red EL component either above or under the blue EL component. Each EL's orientation (arrow) is assigned through the right-hand rule \cite{elchain}. For the red EL in the left case, the braid invariants at the two ends are the same because the two paths are equivalent by smoothly sliding along the red EL. For the right case, their braid invariants are conjugate by the blue EL: $b_1'=b_2^{-1}b_1b_2$ when the blue EL lies above the red EL \cite{SM}. The NACR implies that if the two ELs do not commute $b_1b_2\neq b_2b_1$, one configuration cannot morph into another: noticing that no EL crosses the two end paths during the deformations (inside the black box), and the braid invariants should stay intact.

Further, two noncommutative ELs cannot form a Hopf link, as depicted in Fig. \ref{fig3}(b). One can check that the braid invariants along the central and faraway paths are not identical. It contradicts the static NACR as the two paths are equivalent. An alternative viewpoint from the dynamical NACR starts from two Weyl points (of a Hermitian system) of two  adjacent band gaps separated in the parameter space. By adding gain or loss, two unlinked ELs are spawned from the two Weyl points. The formation of the Hopf link necessitates the illegal crossings in Fig. \ref{fig3}(a). Similarly, we can exclude many other no-go EL structures solely from the NACR without sophisticated model calculations.

We proceed to illustrate a permissible evolution \textcircled{1}$\rightarrow$\textcircled{5} as per the NACR [Fig. \ref{fig3}(c)]. The overall process effectively changes EL configurations from a direct crossing to a ``tangled" crossing. We take the cavity model (\ref{H2D}) yet with a different dynamical protocol:
\begin{equation}\label{H3D}
H=\left(
\begin{array}{ccc}
\sqrt{2}i[\gamma(t)-i\delta] & -\kappa & 0 \\
-\kappa & i[-\tilde{\gamma}+i\tilde{\delta}(\beta)] & -\kappa \\
0 & -\kappa & -\sqrt{2}i[\gamma(t)-i\delta] \\
\end{array}
\right).
\end{equation}
We set $\tilde{\delta}(\beta)=0.3[(\beta-1)^3+3(\beta-1)^2-2]$ ($\beta\in\mathbb{R}$) and slowly vary the parameters $\gamma(t)=t$ to examine the evolution of ELs in the 3D $(\beta,\tilde{\gamma},\delta)$ space at different time instants \cite{SM}. Starting from two non-commutative ELs (red and blue) in \textcircled{1}, we observe their subsequent touching in \textcircled{2}, recombination into staggered rings in \textcircled{3}, and further touching and recombination into tangled ELs in \textcircled{4}\textcircled{5}. The commutative (noncommutative) components are marked in the same (different) color. In all steps, the braid invariants for the representative path $l_1$ stay unchanged as required by the NACR. The touching in \textcircled{2} is through higher-order EPs, while in \textcircled{4}, it leads to the reconnection of ELs with adjusted orientations. Unlike the Hopf link in Fig. \ref{fig3}(b), the configuration in \textcircled{3} has additional EL components at the two wings and is allowed by the NACR. In \textcircled{3}, the first (or third) and second components (counted from left to right) do not commute and cannot be trivially untied. This is verified by the nontrivial braiding of eigenvalue strands in Fig. \ref{fig3}(d) (top panel). The braid invariant for path $l_2$ at step \textcircled{1}\textcircled{5} is obviously trivial. The NACR indicates that the braid invariant for $l_2$ in step \textcircled{3} is also trivial, as verified from the trivial braiding of eigenvalue strands in Fig. \ref{fig3}(d) (bottom panel). Thus the second and fourth (or the first and third) components commute (in the same color). We leave detailed model calculations and more interesting examples of admissible ELs to the Supplemental Material \cite{SM}.

{\color{blue}\textit{Discussions.—}}To conclude, we have demonstrated that the collective behaviors of multiple EPs or ELs in generic multiband non-Hermitian systems are governed by the universal NACR. From this rule, we have uncovered the exotic non-Abelian features of exceptional degeneracies, including the path-dependent annihilation (coalescence) of EPs and the admissible or no-go EL structures. We have further proposed the realizations of these counterintuitive phenomena in acoustic-cavity experiments.

The collective behaviors of multiple EPs or ELs can only be fully captured by the braid invariant which records all the necessary information of non-Abelian topology in multiband non-Hermitian systems. It avoids the oversimplification or ambiguity of the discriminant number or permutation group \cite{pg1,pg3,SM} (a finite subgroup of $B_N$). For instance, a closed path enclosing two EPs of the same topological charge (which cannot annihilate) has trivial band permutations. As a bonus, our framework, in an intuitive and exact way, solves the starting-point problem in stroboscopic encircling multiple EPs \cite{pg2,pg5}. Different from the homotopy-knot theory of separable bands \cite{braidclass1,braidclass2,Hu1}, or an isolated EP \cite{Hu2,EPexp} without a base point, the based path is necessary to account for the interplay of multiple EPs. Choosing another base point ends up with a conjugate braid invariant \cite{allanhatcher}. Yet the non-Abelian physics does not rely on any specific choice.

Applying our results to the 2D or 3D momentum space, the NACR brings distinct non-Abelian features to multiband non-Hermitian metals with exceptional band touchings. It is worth mentioning the key differences from multiband Hermitian topological metals protected by $\mathcal{P}\mathcal{T}$ or $\mathcal{C}_2\mathcal{T}$ symmetry described by quaternion charges \cite{Wu,Bouhon,Tiwari,Bouhonr} (a finite group). There, the non-Abelian topology is attributed to the frame rotations of wave functions, and there is a definite meaning for band gaps and labeling. [Note the subtlety for Floquet systems \cite{Slagerr}.] In stark contrast, the complex eigenvalues and defective degeneracies in non-Hermitian settings invalidate a globally consistent numbering of energy bands and band gaps. The non-Abelian topology is encoded in the eigenenergies. Furthermore, the (second-order) EPs (ELs) are defective and stable without symmetry requirements.

Besides acoustic cavities, the illustrated models and phenomena could also be realized and observed in other platforms like coupled optical waveguides \cite{waveguide1,waveguide2,waveguide3,encircleEP5,encircleEP7} or ring resonators \cite{ringcavity1,ringcavity2,ringcavity3,ringcavity4,fansh1,fansh2}. Besides the EP annihilation (coalescence) and admissible EL structures presented here, the NACR can be utilized to analyze various other collective phenomena, e.g., the exchanges or braidings of EPs or ELs, where the infinite many braid-group elements should give rise to unique non-Abelian properties. Our findings are generic with far-reaching implications in various fields, including optics and photonics to microwaves and acoustics. They should motivate further research on the applications and functionality based on exceptional non-Hermitian physics. 

\begin{acknowledgments}
This work is supported by the National Key Research and Development Program of China (Grant No. 2022YFA1405800), the NSFC under Grants No. 11974413 and No. T2121001, the Strategic Priority Research Program of Chinese Academy of Sciences under Grant No. XDB33000000, and the start-up grant of IOP-CAS (H. H.). K. D. is supported by the NSFC under Grant No. 12174072 and Natural Science Foundation of Shanghai (No. 21ZR1403700).
\end{acknowledgments}

\newpage
\onecolumngrid
\renewcommand{\theequation}{S\arabic{equation}}
\renewcommand{\thefigure}{S\arabic{figure}}
\renewcommand{\thetable}{S\arabic{table}}
\setcounter{equation}{0}
\setcounter{figure}{0}
\setcounter{table}{0}

\begin{center}
    {\bf \large Supplementary Material for ``Exceptional Non-Abelian Topology in Multiband Non-Hermitian Systems"}
\end{center}

\noindent This supplementary material provides details on:

\noindent (I) the derivation of the homotopy invariants for closed paths in the parameter space; \\
\noindent (II) the relation between the static non-Abelian conservation rule (NACR) and other non-Hermitian theorems;\\
\noindent (III) the proof of the dynamical NACR with varying system parameters;\\
\noindent (IV) the merging of two EPs with opposite charges in the two-band case; \\
\noindent (V) the derivation of braid invariants at initial and final stages; \\
\noindent (VI) theoretical design of model parameters;\\
\noindent (VII) the derivation of the EP trajectories for model Eq. (4) in the main text; \\
\noindent (VIII) the conjugate relation; \\
\noindent (IX) the model realizations of various EL configurations of Fig. 3 in acoustic cavities; \\
\noindent (X) the example of the EL emergence allowed by the conservation rule;\\
\noindent (XI) comparison between different topological invariants.

\subsection{(I) The derivation of the homotopy invariants for closed paths in the parameter space}
Our formalism starts with the $N\times N$ non-Hermitian Hamiltonian $H(\bm{\lambda})=H(\lambda_1,\lambda_2,...,\lambda_d)$ with $d$ the dimension of parameter space. $\lambda_1,\lambda_2,...,\lambda_d$ are system paramters. In the general case, the parameter space is punctured by some exceptional degeneracies, at which the Hamiltonians become defective and both the eigenenergies and eigenvectors coalesce. Exceptional points (EPs) of second order are stable in the two-dimensional (2D) parameter space and extend to exceptional lines (ELs) in 3D. For either case, we consider the non-Hermitian topology in the punctured parameter space by focusing on the closed loops enclosing the degeneracies as depicted in Fig. 1a of the main text. For a chosen path $\Gamma$, the non-Hermitian Hamiltonian $H(\bm{\lambda})$ maps path $\Gamma$ of the parameter space to a closed loop in the space of the $N\times N$ matrix $X_N$ (classifying space). We resort to the fundamental group (i.e., the first-order homotopy group) to classify these closed paths in $X_N$ \cite{allanhatcher}. We denote the eigenvalues and eigenvectors of $H(\bm{\lambda})$ as
\begin{eqnarray}
H(\bm{\lambda})|\Psi_n(\bm{\lambda})\rangle=E_n(\bm{\lambda})|\Psi_n(\bm{\lambda})\rangle,~~~n=1,2,...,N.
\end{eqnarray}

To obtain $X_N$, we note that (1) the Hamiltonian $H(\bm{\lambda})$ is fully determined by its eigenvalues and eigenvectors; (2) the Hamiltonian $H(\bm{\lambda})$ on path $\Gamma$ possesses separable bands \cite{Fuliang}, i.e., $E_i\neq E_j$ for any $i\neq j$ on the path $\Gamma$. The eigenvectors $|\Psi_n(\bm{\lambda})\rangle$ (n=1,2,...,N) are linearly independent. Hence $X_N$ can be represented as \cite{braidclass1,braidclass2,Hu1,Hu2,FSH}
\begin{eqnarray}
X_N=\frac{\textrm{Conf}_N(\mathbb{C})\times(\textrm{GL}(N)/\textrm{GL}(1)^N)}{S_N},
\end{eqnarray}
where the eigenvalue part is characterized by the configuration space of ordered N-tuples $\textrm{Conf}_N(\mathbb{C})$. The eigenvector part is described by $\textrm{GL}(N)/\textrm{GL}(1)^N$. Here $\textrm{GL}(N)$ is the general linear group of degree $N$. Since $|\Psi_n(\bm{\lambda})\rangle$ and $c|\Psi_n(\bm{\lambda})\rangle~(c\neq 0)$ represents the same right eigenvector subject to the biorthogonal normalization \cite{binormal}, the gauge degree of freedom should be moved out. The additional factor $S_N$ (symmetric group of degree $N$) comes from the redundancy of simultaneous permutations of eigenvalues and their corresponding eigenvectors. The fundamental group of $X_N$ based at a point $\bm{P}$ in the parameter space can be calculated as \cite{braidclass1,braidclass2,Hu1,Hu2}
\begin{eqnarray}
\pi_{1}(\bm{P},X_N)=B_N.
\end{eqnarray}
To derive the above equation, we have used the mathematical results: (1) $\pi_1(\bm{P},\textrm{Conf}_N(\mathbb{C})/S_N)=B_N$; (2) $\pi_1(\bm{P},\textrm{U}(N)/\textrm{U}(1)^N)=0$; (3) the two spaces $\textrm{U}(N)/\textit{U}(1)^N$ and $\textrm{GL}(N)/\textrm{GL}(1)^N$ are homotopy equivalent. $B_N$ is the braid group. $N=2$ corresponds to the Abelian case with $B_2=\mathbb{Z}$. Otherwise, $B_N$ is non-Abelian. The above exceptional band theory indicates that each path in the parameter space is assigned a braid-valued topological invariant. It dictates how the complex eigenenergies evolve along the path and can be intuitively represented as a braid diagram. The braid invariant provides a strict and complete description for multiband systems with multiple exceptional degeneracies. Notice that the homotopy invariants discussed in the main text are based on a selected base point $\bm{P}$, while the homotopy invariants used in previous papers \cite{braidclass1,braidclass2,Hu1,Hu2} are non-based. They are related by the so-called conjugacy class \cite{allanhatcher} of the braid group. The based homotopy invariant is necessary to analyse the collective behaviors involving multiple exceptional degeneracies.

\subsection{(II) The relation between the static non-Abelian conservation rule (NACR) and other theorems}

The static NACR for non-Hermitian Hamiltonian dictates that two topologically equivalent paths have identical braid invariants. Here we prove the non-Hermitian no-go theorem \cite{Hu2}, the non-Hermitian doubling theorem \cite{Yang} and the flow conservation \cite{elchain} based on this static NACR.

\subsubsection{(a) The relation between the static NACR and the non-Hermitian no-go theorem}
\begin{figure*}[btp]
\includegraphics[width=.7\textwidth]{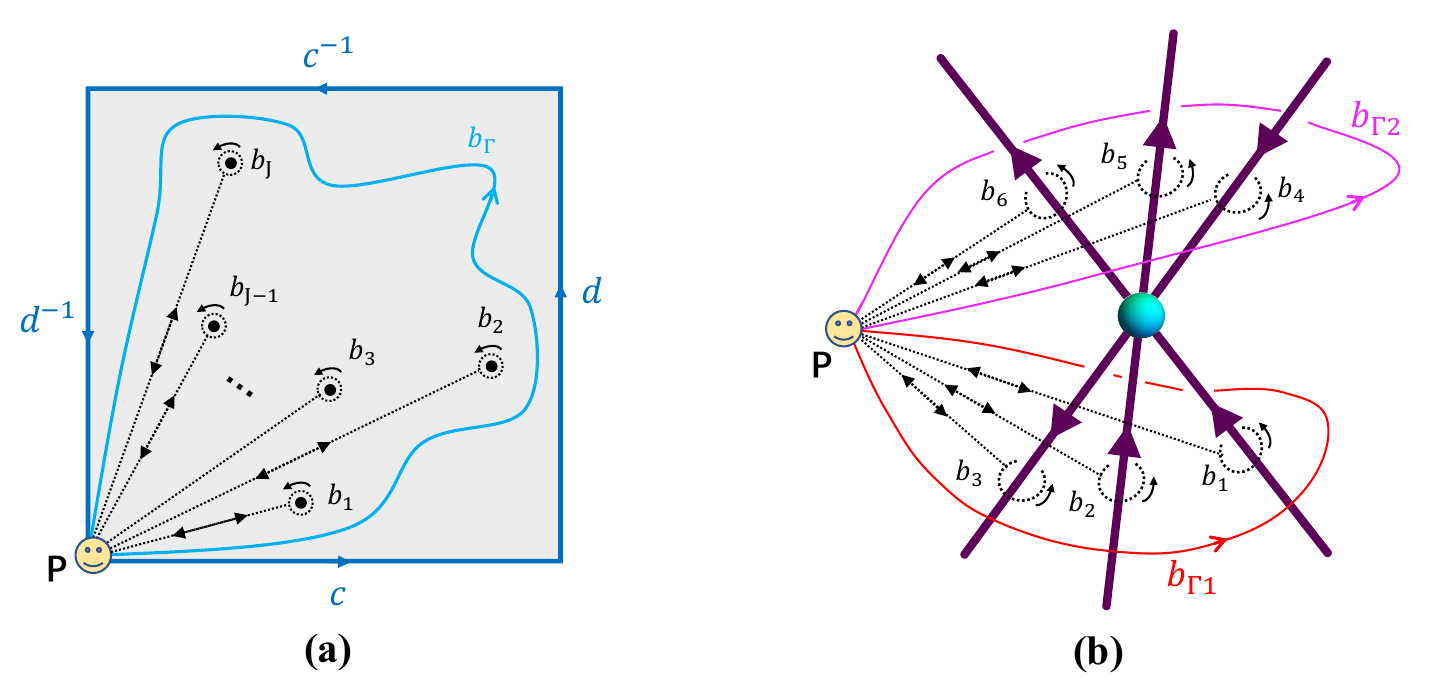}
\caption{(a) Sketch of the proof of the non-Hermitian no-go theorem and doubling theorem on a 2D periodic lattice. (b) Sketch of the proof of the flow conservation at an exceptional junction (cyan dot) for 3D non-Hermitian systems. The purple arrow marks the flow of the EL.}
\label{fig_SMII}
\end{figure*}

The non-Hermitian no-go theorem \cite{Hu2} deals with the EPs on a 2D periodic lattice. It tells that the multiplication of the braid invariants of all the EPs (without any base point) in the 2D Brillouin zone (BZ) is inside the commutator subgroup $[B_N, B_N]=\{ uvu^{-1}v^{-1}|(u,v)\in B_N \}$. Here, we prove this theorem.

According to the exceptional band theory, the $j$th EP ($j=1,2...,J$) is assigned a braid-valued topological invariant $b_j$ along the corresponding path based at the point $\bm{P}$, where $J$ is the number of total EPs in the BZ. For convenience, we choose the base point $\bm{P}$ at the BZ corner, as sketched in Fig. \ref{fig_SMII}(a). For any other base point, we can always shift it to the corner without affecting the conclusion. It is easy to see the braid invariant along the path $\Gamma$ enclosing all the EPs based at $\bm{P}$ is $b_1b_2...b_{J-1}b_{J}$. The path $\Gamma$ can be deformed to the BZ boundary without touching any EPs. According to the static NACR, $b_1b_2...b_{J-1}b_{J}=cdc^{-1}d^{-1}\in[B_N,B_N]$, with $c,d,c^{-1},d^{-1}$ the braid invariants along the four edges of BZ. Further note that the commutator subgroup has the properties for any $x, y\in\textrm{B}_n$: (1) $x[u,v]x^{-1}=[xux^{-1},xvx^{-1}]\in[\textrm{B}_n,\textrm{B}_n]$; (2) $yb_1y^{-1}b_2...b_J=(yb_1y^{-1}b_1^{-1})(b_1b_2...b_J)=[y,b_1][u,v]\in[\textrm{B}_n,\textrm{B}_n]$; (3) $b_{i+1}b_i=b_i^{-1}(b_ib_{i+1})b_i$. These properties ensure that the ordering of the EPs (i.e., labelling of EPs from $1$ to $J$) and taking the conjugate element of the element $b_j$ do not affect the results. Hence the multiplication of the non-based braid invariants is still inside the commutator subgroup.

\subsubsection{(b) The relation between the static NACR and the non-Hermitian doubling theorem}
The non-Hermitian doubling theorem \cite{Yang} tells that the second-order EPs must come in pairs on a 2D non-Hermitian periodic lattice. In fact, the doubling theorem is a direct consequence of the non-Hermitian no-go theorem by noting that the discriminant number along path $\Gamma$ is equal to the writhe of the braid invariant $\nu_\Gamma$, i.e., $\nu_\Gamma=n_+(\Gamma)-n_-(\Gamma)$ \cite{Hu2}, where $n_{\pm}$ denotes the number of over/under crossings of $\nu_\Gamma$, respectively. For example, the discriminant number is $2$ for the braid invariant $\tau_1^{-1}\tau_2\tau_1\tau_3$ with $n_+=3$ and $n_-=1$.  From the discussions in the last subsubsection, the braid invariant along the four edges of the 2D BZ lies inside the commutator subgroup $[B_N,B_N]=\{ uvu^{-1}v^{-1}|(u,v)\in B_N \}$ with zero discriminant number. Thus the sum of the discriminant number of all EPs in the 2D BZ vanishes. The second-order EPs (with discriminant number $\pm 1$) must come in pairs on a 2D periodic lattice.

\subsubsection{(c) The relation between the static NACR and the flow conservation}
The flow conservation \cite{elchain} tells that the number of in-flowing ELs and out-flowing ELs in 3D non-Hermitian system must be equal when several oriented ELs (composed of second-order EPs) meet at an exceptional junction. Here we prove the theorem using the static NACR.

Let us consider multiple ELs (e.g., six, as shown in Fig. \ref{fig_SMII}(b)) composed of second-order EPs that meet at a junction. According to the exceptional band theory, the $j$th EL ($j=1,2...,6$) is assigned a braid-valued topological invariant $b_j$ along the path with base point $\bm{P}$, and the corresponding discriminant number is labeled as $\nu(b_j)$. Each EL is also assigned a direction, or flow \cite{elchain} by requiring a small loop enclosing the EL (through the right-hand rule) to possess a positive discriminant number. If $\nu(b_1/b_2/b_3)>0$ (or $\nu(b_1/b_2/b_3)<0$), the corresponding EL in-flows (out-flows) the junction, while the corresponding EL out-flows (in-flows) the junction for $\nu(b_4/b_5/b_6)>0$ (or $\nu(b_4/b_5/b_6)<0$). It is easy to see the braid invariant along path $\Gamma_1$ is $b_{\Gamma_1}=b_1b_2b_3$ with discriminant number $\nu_{\Gamma_1}=\nu(b_1)+\nu(b_2)+\nu(b_3)=n_{\rm{in}}(\Gamma_1)-n_{\rm{out}}(\Gamma_1)$, where $n_{\rm{in}}(\Gamma_1)$ (or $n_{\rm{out}}(\Gamma_1)$) is the number of ELs inside the path $\Gamma_1$ in-flowing (or out-flowing) the node. Similarly, the braid invariant along path $\Gamma_2$ is $b_{\Gamma_2}=b_4b_5b_6$ with  discriminant number $\nu_{\Gamma_2}=\nu(b_4)+\nu(b_5)+\nu(b_6)=n_{\rm{out}}(\Gamma_2)-n_{\rm{in}}(\Gamma_2)$, where $n_{\rm{in}}(\Gamma_2)$ (or $n_{\rm{out}}(\Gamma_2)$) is the number of ELs inside the path $\Gamma_2$ in-flowing (or out-flowing) the node. Obviously the path $\Gamma_1$ can be continuously deformed into $\Gamma_2$ without touching any ELs. According to the static NACR, the braid invariants along these two paths are equal $b_{\Gamma_1}=b_{\Gamma_2}$. Since the discriminant number is fully determined by the braid invariant (i.e., its writhe), we have
\begin{eqnarray}
n_{\rm{in}}(\Gamma_1)+n_{\rm{in}}(\Gamma_2)=n_{\rm{out}}(\Gamma_1)+n_{\rm{out}}(\Gamma_2).
\end{eqnarray}
This is exactly the flow conservation which states that the numbers of in-flowing and out-flowing ELs at the junction must be equal.

\subsection{(III) The proof of the dynamical non-Abelian conservation rule (NACR) with varying system parameters}

The dynamical NACR governs the motions of EPs or ELs in 2D or 3D parameter space with varying system parameters. In this section, we prove the dynamical NACR.
\begin{figure*}[!thb]
\includegraphics[width=.5\textwidth]{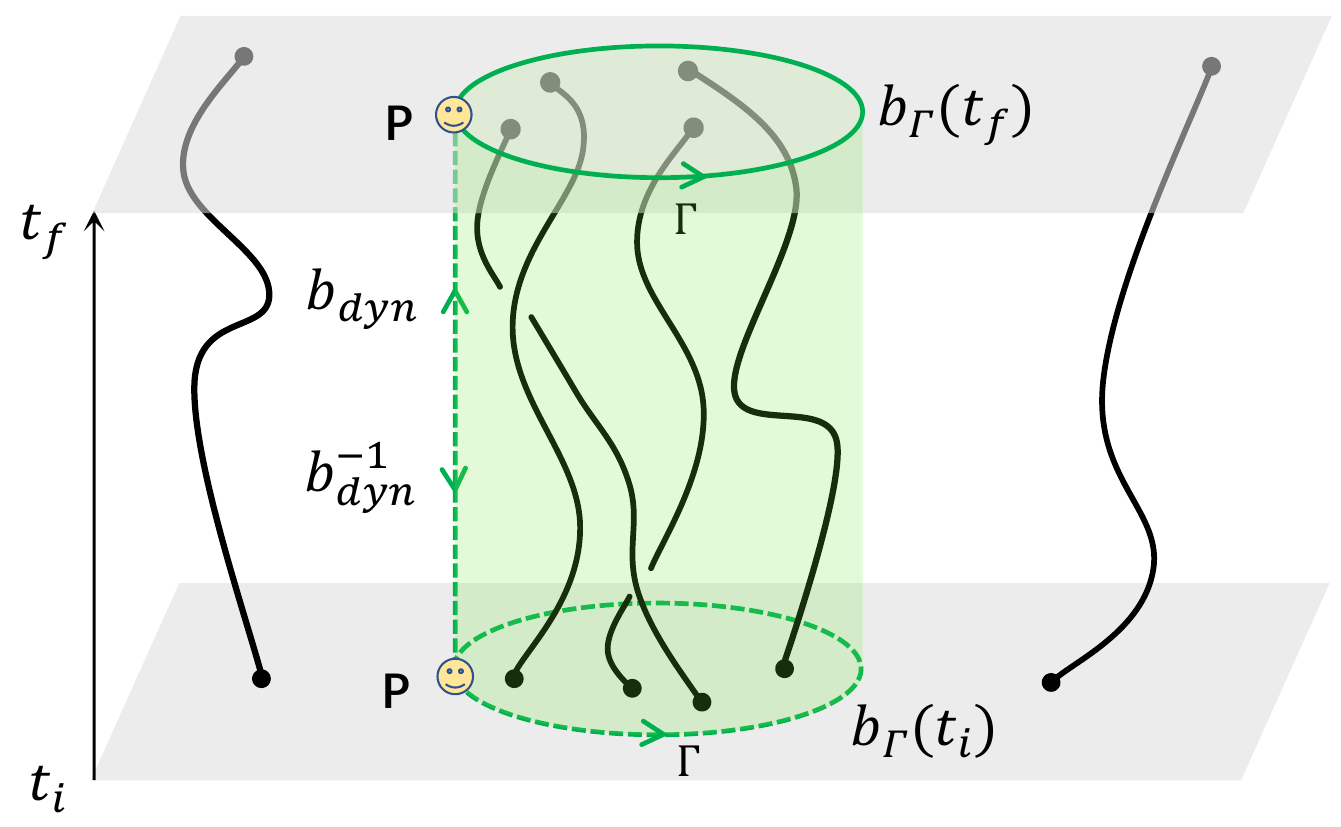}
\caption{Sketch of the proof of the dynamical NACR. The black lines are the trajectories of the EPs in the 2D parameter space with varying parameters from $t=t_i$ to $t=t_f$. During the whole process, there is no EP entering or escaping the path $\Gamma$, i.e., no EP trajectories touching the green cylinder surface.}
\label{fig_SMIII}
\end{figure*}

We focus on the case of EPs in the 2D parameter space. The same arguments apply to the ELs in 3D. Let us consider a fixed path $\Gamma$ with base point $\bm{P}$. The braid invariants along the path $\Gamma$ at the initial and final stages are denoted as $b_{\Gamma}(t_i)$ and $b_{\Gamma}(t_f)$, respectively. Figure \ref{fig_SMIII} schematically shows the motions of the EPs in the parameter space with the stroboscopic variation of system parameters starting from $t=t_i$ to $t=t_f$. The configurations of the EPs in the 2D parameter space may change, yet there is no EP crossing the path $\Gamma$, i.e., no trajectories of EPs enter or escape the cylinder surface (green) in Fig. \ref{fig_SMIII}. In the 3D space spanned by the time dimension and the original 2D parameters, the path $\Gamma$ at $t_f$ (solid green line) can be smoothly deformed into the concatenated path which goes from the base point $\bm{P}$ at $t=t_f$ to $\bm{P}$ at $t=t_i$, and then follows $\Gamma$ at $t=t_i$ and finally goes back to $\bm{P}$ at $t=t_f$. According to the static NACR in the 3D parameter space, the braid invariants along these two topologically equivalent paths are equal. By denoting the eigenvalue braiding along the path element starting from $\bm{P}$ at $t=t_i$ to $\bm{P}$ at $t=t_f$ as $b_{dyn}$, we have
\begin{eqnarray}\label{dcrs}
b_{\Gamma}(t_f)=b_{dyn}^{-1}~b_{\Gamma}(t_i)~b_{dyn}.
\end{eqnarray}
Here $b_{dyn}$ can be regarded purely as a dynamical factor describing the accumulated braiding from time $t_i$ to $t_f$ of the instantaneous eigenenergy at the base point $\bm{P}$. Equation (\ref{dcrs}) is nothing but the dynamical NACR. As the dynamical factor $b_{dyn}$ acts indiscriminately on all the closed paths based at $\bm{P}$, it would not affect the physics involving multiple EPs.

\subsection{(IV) The annihilation of two EPs with opposite charges}
\begin{figure*}[!b]
\includegraphics[width=.5\textwidth]{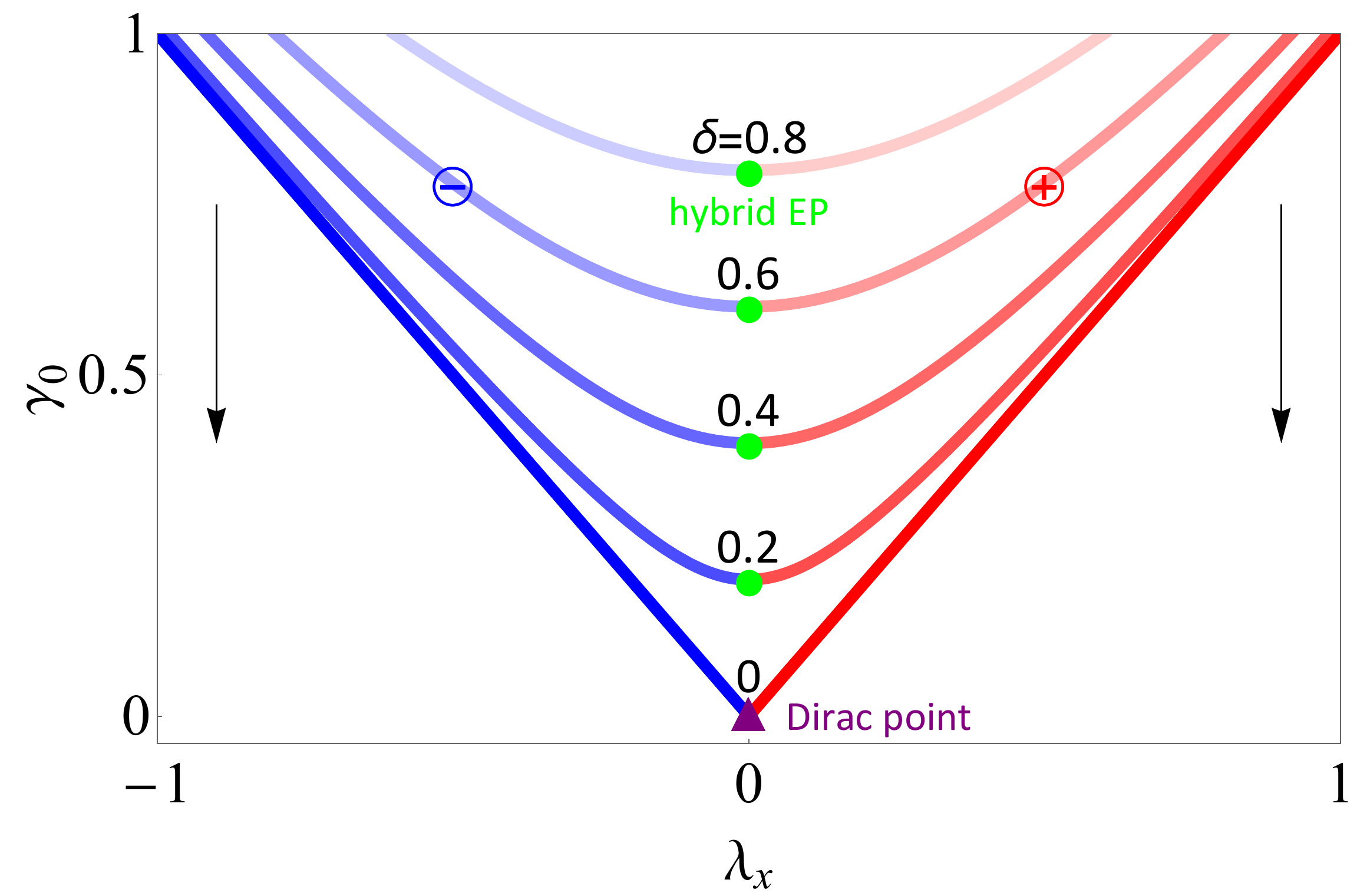}
\caption{Annihilation process of two EPs with opposite charges in model (\ref{2bandmodel}) by slowly decreasing the parameter $\gamma_0$ (black arrow). The red/blue lines are the trajectories of EP with positive/negative charge, respectively. From bottom to top, $\delta_0=0, 0.2, 0.4, 0.6, 0.8$. The two EPs either merge into a Dirac point (purple triangle) at $\gamma_0=\delta_0=0$ or a hybrid EP (green dot) at $\gamma_0=\delta_0\neq 0$.}
\label{fig_SMIV}
\end{figure*}
In the main text, we have discussed the annihilation of two EPs for generic multi-band non-Hermitian systems. In the two-band case, the braid invariant reduces to the discriminant number or topological charge since $B_2=\mathbb{Z}$. Two EPs of opposite charges annihilate each other, which can be simply written as $1+(-1)=0$. To be more clear, we examine the fate of two EPs with opposite charges when brought together through the following two-band model:
\begin{equation}\label{2bandmodel}
H=\lambda_x\sigma_x+(\lambda_y+i\gamma_0)\sigma_y+\delta_0 \sigma_z,
\end{equation}
where $\lambda_x,\lambda_y, \gamma_0$, $0\leq\delta_0<1$ are real parameters, and $\sigma_{x,y,z}$ are the Pauli matrices. The eigenspectra are
\begin{equation}
E=\pm\sqrt{\lambda_x^2+\lambda_y^2+\delta_0^2-\gamma_0^2+2i\gamma\lambda_y}.
\end{equation}

It is easy to see that there exist a pair of EPs located at $(\lambda_x,\lambda_y)=(\pm\sqrt{\gamma_0^2-\delta_0^2},0)$ with opposite charges. The discriminant number is $\nu_{\Gamma}=1$ for the EP with $\lambda_x>0$ and $\nu_{\Gamma}=-1$ for the EP with $\lambda_x<0$. Here $\Gamma$ is a small path enclosing the corresponding EP. When we slowly deduce $\gamma_0$ from $1$ to $0$ at fixed $\delta_0$, the trajectories of the two EPs are plotted in Fig. \ref{fig_SMIV}. They first approach each other (when $\gamma_0>\delta_0$), then merge (at $\gamma_0=\delta_0$) and annihilate (when $\gamma_0<\delta_0$). There are two different cases. (1) If $\delta_0=0$, the two EPs merge into a transient Dirac point at $(\lambda_x,\lambda_y)=(0,0)$. This is the inverse process of the well-known splitting of a Dirac point into two opposite EPs \cite{dptoep1,dptoep2} as observed in photonic crystals with radiation loss \cite{epscience}. The Dirac point is unstable in 2D without symmetry protection and will be gapped out under perturbation. (2) For $\delta_0\neq0$, the two EPs merge into the so-called hybrid EP \cite{Fuliang,hep1,hep2,hep3}, a defective singularity with anisotropic dispersion, at which we have $\nu_{\Gamma}=0$. The system is fully gapped with further decreasing $\gamma_0$. In both cases, the two original EPs with opposite charges annihilate.

\subsection{(V) The derivation of braid invariants at initial and final stages}
\begin{figure*}[!b]
\includegraphics[width=.4\textwidth]{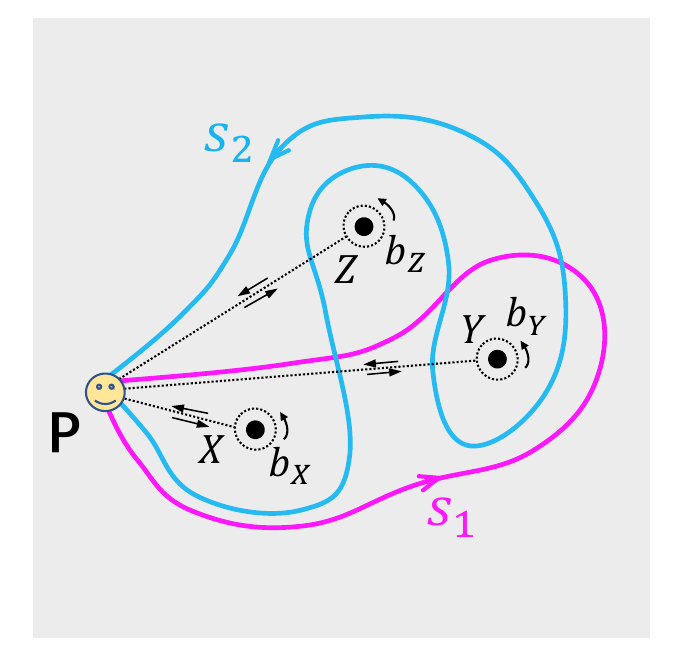}
\caption{Analysis of annihilation and coalescence of EPs. Two paths $S_1$ (magenta) and $S_2$ (cyan) starting from $\bm{P}$ enclose two EPs located at $X$ (blue) and $Y$ (red) in the 2D parameter space. In between there is another EP located at $Z$. The annihilation or coalescence of the two EPs is generally path-dependent and dictated by the braid invariant of the enclosing path.}
\label{fig_SMV}
\end{figure*}

As illustrated in Fig. 2(a) of the main text, the braid invariant at the initial stage $b_{\Gamma}(t_i)$ and final stage $b_{\Gamma}(t_f)$ can be different according to the NACR. Here, we explain this difference in detail. Suppose $X$ and $Y$ were initially created pairwise from a Dirac point or a hybrid EP at $t_i$. We have $b_Y(t_i)=b_X^{-1}(t_i)$ and $b_{\Gamma}(t_i)=1$. As depicted in Fig. \ref{fig_SMV}, $b_{X/Y/Z}$ are the braidings (based at $\rm{P}$) along the local path encircling $X/Y/Z$. With time-varying parameters, the two EPs $X$ and $Y$ do not annihilate each other at the final stage $t_f$ except $b_X(t_i)$ and $b_Z(t_i)$ commute, i.e., $b_X(t_i)b_Z(t_i)=b_Z(t_i)b_X(t_i)$.

We note that the path $\Gamma$ at $t_i$ and $t_f$ is smoothly deformed to topologically distinct path $s_1$ and $s_2$ at some intermediate moment $t_m$, as depicted in Fig. \ref{fig_SMV}. [See also Fig. 2(a) in the main text.] The braiding along $s_1$ and $s_2$ at $t_m$ can be obtained as
\begin{eqnarray}\label{bs1s2tm}
b_{s_1}(t_m)=b_X(t_m)b_Y(t_m),~~~b_{s_2}(t_m)=b_X(t_m)b_Z^{-1}(t_m)b_Y(t_m)b_Z(t_m).
\end{eqnarray}
According to the dynamical NACR, we have
\begin{eqnarray}\label{dy}
b_{X/Y/Z}(t_m)=b_{dyn}^{-1}(t_i\rightarrow t_m)b_{X/Y/Z}(t_i)b_{dyn}(t_i\rightarrow t_m).
\end{eqnarray}
Here $b_{dyn}(t_i\rightarrow t_m)$ is the dynamical factor accounting for the accumulated braiding of (instantaneous) eigenenergy from time $t_i$ to $t_m$ at the base point $\bm{P}$. Substituting Eq. (\ref{dy}) into Eq. (\ref{bs1s2tm}), we have
\begin{eqnarray}\label{bs1s2tm2}
\begin{split}
&b_{s_1}(t_m)=b_{dyn}^{-1}(t_i\rightarrow t_m)b_X(t_i)b_Y(t_i)b_{dyn}(t_i\rightarrow t_m)=1,\\
&b_{s_2}(t_m)=b_{dyn}^{-1}(t_i\rightarrow t_m)b_X(t_i)b_Z^{-1}(t_i)b_X^{-1}(t_i)b_Z(t_i)b_{dyn}(t_i\rightarrow t_m).
\end{split}
\end{eqnarray}
Further, the path $\Gamma$ at the final stage is topologically equivalent to the path $s_2$ at $t_m$. According to dynamical NACR we have
\begin{eqnarray}\label{bgmf1}
b_{\Gamma}(t_f)=b_{dyn}^{-1}(t_m\rightarrow t_f)b_{s_2}(t_m)b_{dyn}(t_m\rightarrow t_f).
\end{eqnarray}
Inserting Eq. (\ref{bs1s2tm2}), Eq. (\ref{bgmf1}) is recast into
\begin{eqnarray}\label{bgmf2}
b_{\Gamma}(t_f)=b_{dyn}^{-1}(t_i\rightarrow t_f)b_X(t_i)b_Z^{-1}(t_i)b_X^{-1}(t_i)b_Z(t_i)b_{dyn}(t_i\rightarrow t_f),
\end{eqnarray}
here we have used $b_{dyn}(t_i\rightarrow t_f)=b_{dyn}(t_i\rightarrow t_m) b_{dyn}(t_m\rightarrow t_f)$ and $b_{dyn}(t_i\rightarrow t_f)^{-1}=b_{dyn}(t_m\rightarrow t_f)^{-1}b_{dyn}(t_i\rightarrow t_m)^{-1}$. After appropriately choosing base point $P$ such that $b_{dyn}(t_i\rightarrow t_f)=1$ holds, we have the final result:
\begin{eqnarray}\label{Btf}
b_{\Gamma}(t_f)=b_X(t_i)b_Z^{-1}(t_i)b_X^{-1}(t_i)b_Z(t_i).
\end{eqnarray}
If initially $b_X(t_i)$ and $b_Z(t_i)$ commute, i.e., $b_X(t_i)b_Z(t_i)=b_Z(t_i)b_X(t_i)$, $b_{\Gamma}(t_f)=1$. It indicates that the two EPs $X$ and $Y$ also annihilate each other at the final stage once they meet. On the contrary, for the case of $b_X(t_i)b_Z(t_i)\neq b_Z(t_i)b_X(t_i)$, the two EPs do not annihilate each other due to $b_{\Gamma}(t_f)\neq1$. Without any confusion, we will ignore the time parameter and simplify Eq. (\ref{Btf}) as
\begin{eqnarray}
b_{\Gamma}(t_f)=b_Xb_Z^{-1}b_X^{-1}b_Z.
\end{eqnarray}

\subsection{(VI) Theoretical design of model parameters}
In the main text, we have demonstrated the exceptional non-Abelian topology in Fig. 2 and Fig. 3. The dynamical protocols or the analytic expressions of cavity parameters $\gamma(t)$ and $\tilde{\delta}(t)$ in the main text are selected for illustration purposes only. There are many other choices with the same exceptional physics. Here we focus on the coupled cavity system described by Eq. 4 (in the main text) and outline the key points of choosing the appropriate dynamical protocols.
\begin{figure*}[h]
\includegraphics[width=.78\textwidth]{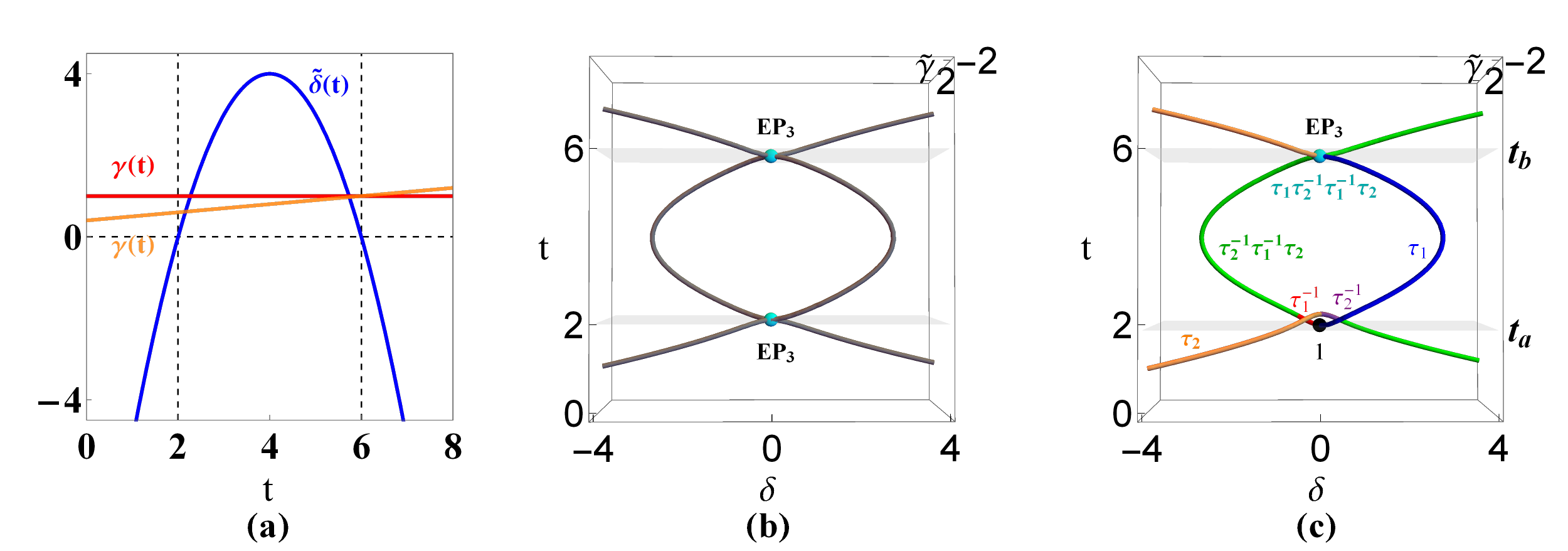}
\caption{Theoretical design of dynamical protocols. (a) The cavity parameter $\tilde{\delta}(t)$ as a quadratic function and $\gamma(t)$ as a constant or linear function. Here $\tilde{\delta}(t)=4-(t-4)^2$ (blue curve), $\gamma(t)=1$ (red line) and $\gamma(t)=0.4+0.1t$ (orange line). (b) The intermediate case. The trajectories of EPs of Hamiltonian Eq. 4 (in the main text) in the 2D $(\delta, \tilde{\gamma})$ space with $\gamma(t)=1$, $\tilde{\delta}(t)=4-(t-4)^2$. (c) The EP loci with $\gamma(t)=0.4+0.1t$, $\tilde{\delta}(t)=4-(t-4)^2$. EPs are marked with colors according to their braidings. The base point $\bm{P}$ is pinned at $(\delta, \tilde{\gamma})=(0, -3)$. The cyan dots represent $\mathrm{EP}_3$.}
\label{fig_SMNew1}
\end{figure*}

We aim to realize the EP loci as sketched in Fig. (\ref{fig_SMNew1})(c). The upper part of the structure has an $\mathrm{EP}_3$ with several $\mathrm{EP}_2$ rays emanating from it. The lower part of the structure consists of two tangled EP loci. To this end, we start from an intermediate case with two $\mathrm{EP}_3$ as depicted in Fig. (\ref{fig_SMNew1})(b) and break the $\mathrm{EP}_3$ at the lower part. Note that the EP loci at the lower part cannot be fully detached, as ensured by the NACR. The recipe of realizing the intermediate structure with two $\mathrm{EP}_3$ in Fig. (\ref{fig_SMNew1})(b) is that the Hamiltonian Eq. 4 (in the main text)  has an $\mathrm{EP}_3$ located at $(\delta, \tilde{\gamma})=(0,0)$ when $\tilde{\delta}(t)=0$, $\gamma(t)=1$. Thus we can take $\tilde{\delta}(t)$ as a quadratic function and $\gamma(t)=1$ as constant. For instance, we set $\tilde{\delta}(t)=4-(t-4)^2$, and the two $\mathrm{EP}_3$ appear at two time instants $t_1=2$, $t_2=6$ in Fig. (\ref{fig_SMNew1})(b). Next, we remove the $\mathrm{EP}_3$ at the lower part $t_1=2$ from this intermediate case by taking $\gamma(t)$ as the linear form, e.g., $\gamma(t)=0.4+0.1t$ as in Fig. (\ref{fig_SMNew1})(a) (the orange line). The $\mathrm{EP}_3$ at $t_2=6$ survives while the $\mathrm{EP}_3$ at $t_1=2$ breaks into several second-order EPs, and the tangle EP loci appears as depicted in Fig. \ref{fig_SMNew1}(c). 

The non-annihilation of two EPs of opposite charges as a manifestation of non-Abelian topology can be verified by the local braid invariant as marked in Fig. \ref{fig_SMNew1}(c). Here the $\mathrm{EP}_3$ at $t_2=6$ has nontrivial braid invariant $\tau_1\tau_2^{-1}\tau_1^{-1}\tau_2$. Besides the analytical expressions above, we have tried other workable dynamical protocols (1) $\gamma(t)=0.2+0.2t$, $\tilde{\delta}(t)=2-2(t-3)^2$; (2) $\gamma(t)=0.25+0.15t$, $\tilde{\delta}(t)=4-(t-3)^2$; (3) $\gamma(t)=0.55+0.1t$, $\tilde{\delta}(t)=3-0.75(t-2.5)^2$. Similarly, we can construct the analytical expressions for cavity parameters in the main text's Hamiltonian Eq. (5) and Fig. 3(c). The key step is to realize the configurations of ELs with three $\mathrm{EP}_3$ in step \textcircled{2}. We can achieve this by taking $\tilde{\delta}(\beta)$ as a cubic function of $\beta$ and $\gamma(t)=1$ as constant. By adjusting the time-dependence of $\gamma(t)$, e.g., $\gamma(t)=t$, the configurations of ELs in the other steps appear. Also, we have tried other analytical expressions of $\tilde{\delta}(\beta)$, such as (1) $\tilde{\delta}(\beta)=\beta^3-\beta$; (2) $\tilde{\delta}(\beta)=\beta^3-3\beta^2+2\beta$; (3) $\tilde{\delta}(\beta)=0.2(\beta^3-3\beta^2-\beta+3)$. We have further verified that these EL configurations obey the NACR.

\subsection{(VII) The derivation of the EP trajectories}
In this section, we give the detailed derivation of EP trajectories of the following three-band model (Eq. (4) in the main text)
\begin{equation}\label{H2D_SM}
H=\left(
\begin{array}{ccc}
\sqrt{2}i[\gamma(t)+i\delta] & -\kappa & 0 \\
-\kappa & i[\tilde{\gamma}+i\tilde{\delta}(t)] & -\kappa \\
0 & -\kappa & -\sqrt{2}i[\gamma(t)+i\delta] \\
\end{array}
\right),
\end{equation}
where $\kappa,\gamma,\tilde{\gamma},\delta,\tilde{\delta}\in\mathbb{R}$ are parameters of the acoustic cavities. We take $\kappa=1$ as the energy unit. We slowly vary the parameters as $\gamma(t)=(t+1)/4$ and $\tilde{\delta}(t)=1-(t-2)^2$ and investigate the motions of EPs in the 2D $(\delta,\tilde{\gamma})$ space. The characteristic polynomial of the Hamiltonian (\ref{H2D_SM}) can be obtained as
\begin{equation}
    f(\delta,\tilde{\gamma})=\mathrm{det}[E-H]=E^3+g_1E^2+g_2E+g_3,
\end{equation}
with
\begin{equation}
\begin{array}{lll}
   g_1=\tilde{\delta}-i\tilde{\gamma},\\
   g_2=2(\gamma^2-\delta^2-1+2i\gamma\delta),\\
   g_3=2(2\gamma\tilde{\gamma}\delta+\gamma^2\tilde{\delta}-\delta^2\tilde{\delta}+i\tilde{\gamma}\delta^2+2i\gamma\delta\tilde{\delta}-i\gamma^2\tilde{\gamma}).
\end{array}
\end{equation}

\begin{figure*}[!h]
\includegraphics[width=.7\textwidth]{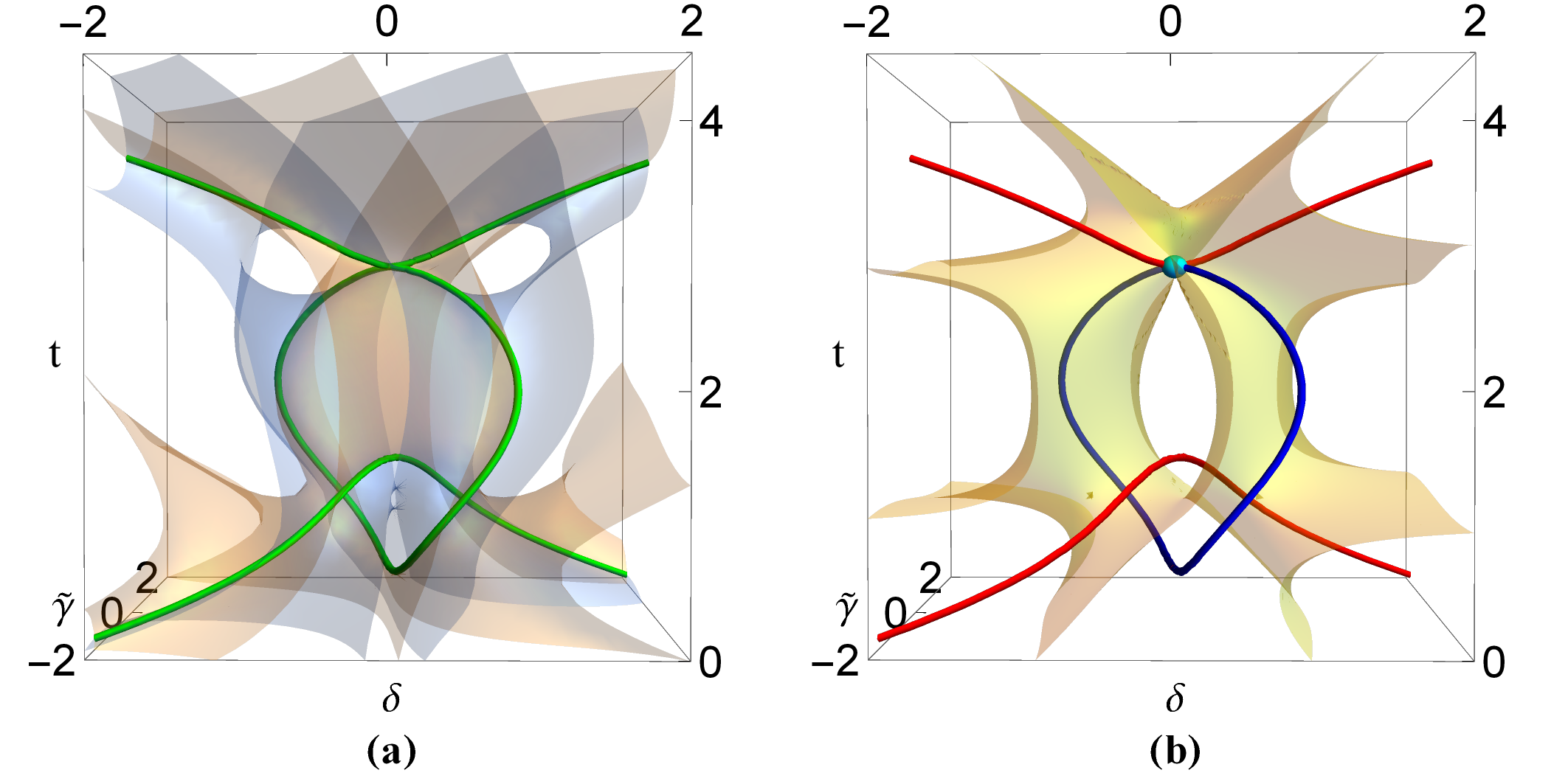}
\caption{The trajectories of EPs of Hamiltonian (\ref{H2D_SM}) in the 2D $(\delta, \tilde{\gamma})$ space with varying parameters. (a) The EP trajectories (green lines) are the intersections of the orange surface with $\mathrm{Re}[\Delta(f)]=0$ and the light blue surface with $\mathrm{Im}[\Delta(f)]=0$. (b) The blue trajectories represent EPs of $E_1=E_2$, while the red trajectories represent EPs $E_2=E_3$. $E_1, E_2, E_3$ are sorted according to their real parts. The two kinds of EPs are separated by the yellow surface of $\mathrm{Re}E_a-\mathrm{Re}E_b=0$. The cyan dot is a third-order EP.}
\label{fig_SMVI}
\end{figure*}

The occurrence of band degeneracies is given by the condition $\Delta(f)=0$. Here $\Delta(f)$ is the discriminant of the characteristic polynomial $f(\delta,\tilde{\gamma})$ and related to the Sylvester matrix:
\begin{equation}
    \Delta(f)=\prod_{i<j}(E_i-E_j)^2=-\mathrm{det}[\mathrm{Sly(g_1,g_2,g_3)}],
\end{equation}
where $E_i$ is the $i$th eigenvalue of $H$. The Sylvester matrix $\mathrm{Sly(g_1,g_2,g_3)}$ explicitly has the following form
\begin{equation}
   \mathrm{Sly(g_1,g_2,g_3)}=\left(
       \begin{array}{ccccc}
         1 & g_1 & g_2 & g_3 & 0 \\
         0 & 1 & g_1 & g_2 & g_3 \\
         3 & 2g_1 & g_2 & 0 & 0 \\
         0 & 3 & 2g_1 & g_2 & 0 \\
         0 & 0 & 3 & 2g_1 & g_2 \\
       \end{array}
     \right).
\end{equation}
With the help of the Sylvester matrix $\mathrm{Sly(g_1,g_2,g_3)}$, the discriminant $\Delta(f)$ can be represented as
\begin{equation}
    \Delta(f)=g_1^2g_2^2-4g_2^3-4g_1^3g_3+18g_1g_2g_3-27g_3^2.
\end{equation}
The complex equation $\Delta(f)=0$ indicates that the following two equations should be satisfied simultaneously:
\begin{eqnarray}
\mathrm{Re}[\Delta(f)]=0, \mathrm{Im}[\Delta(f)]=0.
\end{eqnarray}
Either of which yields a specific set of surfaces in the 3D parameter space $(\delta,\tilde{\gamma}, t)$. Their intersections give the EP trajectories as depicted in Fig. \ref{fig_SMVI}(a).

In non-Hermitian systems, the order of energy levels is not well defined due to the complex energy spectra. In the main text, we sort the eigenvalues $E_i (i=1,2,3)$ in terms of their real parts, $\mathrm{Re}E_1\leqslant \mathrm{Re}E_2\leqslant \mathrm{Re}E_3$. When $\Delta(f)=0$, one of the eigenvalues is $E_a=\frac{g_1g_2-9g_3}{g_1^2-3g_2}-g_1$, and the other two eigenvalues are degenerate at $E_b=-\frac{g_1g_2-9g_3}{2(g_1^2-3g_2)}$.  In Fig. \ref{fig_SMVI}(b), the blue trajectories represent EPs of $E_1=E_2$, while the red trajectories represent EPs of $E_2=E_3$. They are separated by the surface: $\mathrm{Re}E_a-\mathrm{Re}E_b=0$ (the yellow surface in Fig. \ref{fig_SMVI}(b)). In addition, when the condition $g_1=g_2=g_3=0$ is satisfied, all three eigenvalues are degenerate at $E=0$. This condition is satisfied when $\delta=\tilde{\gamma}=0,t=3$. The three eigenvectors coalesce into $[-1,-i\sqrt2,1]^T$. Therefore $(\delta,\tilde{\gamma})=(0,0)$ is a third-order $\mathrm{EP}$ (cyan dot) when $t=3$.

\subsection{(VIII) The conjugate relation}

\begin{figure*}[!h]
\includegraphics[width=.7\textwidth]{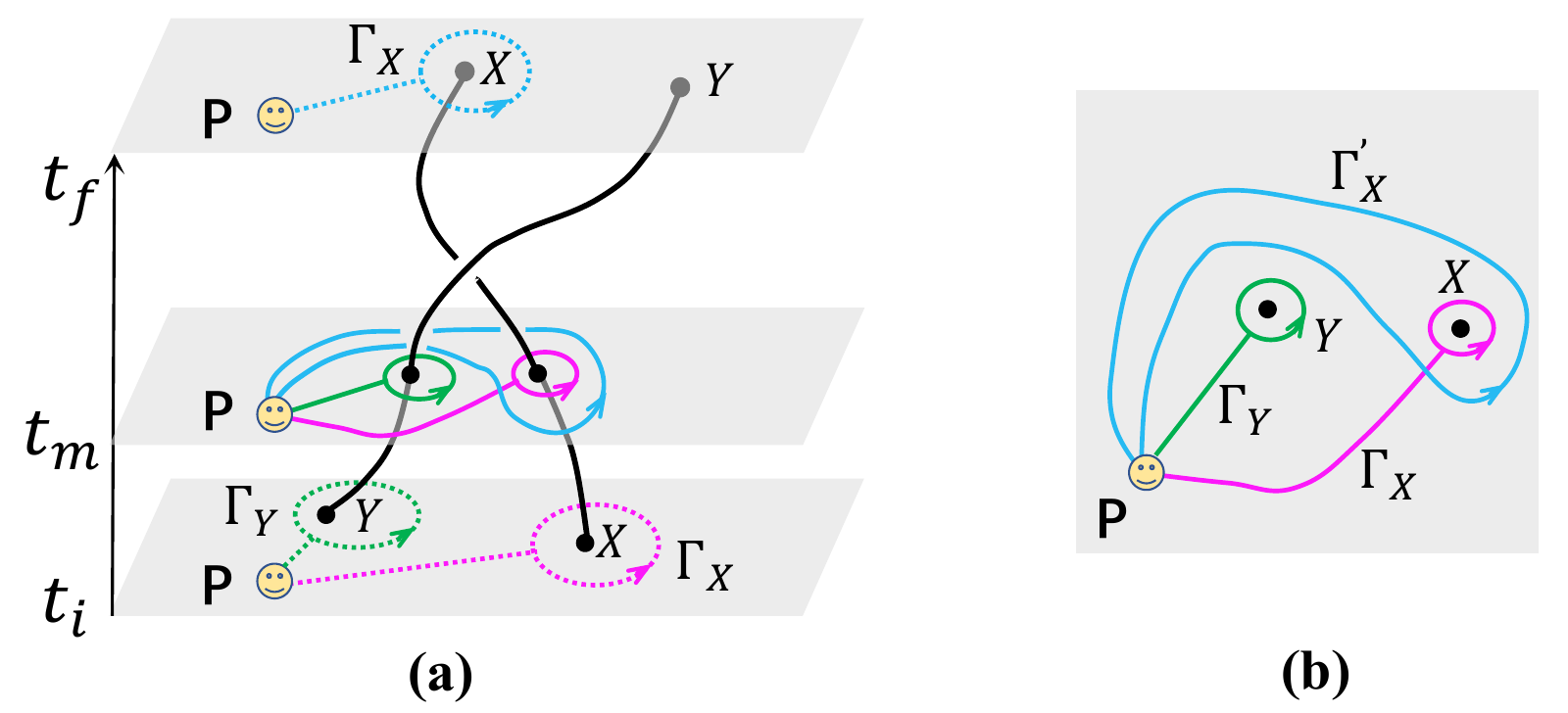}
\caption{Sketch of the proof of the conjugate relation. (a) Time-evolved loci of the two EP $X$, $Y$ in the 3D parameter space (black curves). The target EP ($X$) locus under-crosses another EP ($Y$) locus. (b) Projection of the considered paths at some intermediate time $t_m$ onto the 2D parameter space.}
\label{fig_SMVII}
\end{figure*}

In the main text, we mentioned that if the targeted EP locus under-crosses another EP locus, the braiding of the targeted EP will shift to its conjugate partner. Here we prove this statement. As depicted in Fig. \ref{fig_SMVII}(a), we denote the enclosing path around $X/Y$ as $\Gamma_{X/Y}$ and the corresponding braiding invariant as $b_{\Gamma_{X/Y}}$, respectively. The conjugate relation is represented as $b_{\Gamma_{X}}(t_f)=b_{\Gamma_{Y}}(t_i)^{-1}b_{\Gamma_{X}}(t_i)b_{\Gamma_{Y}}(t_i)$.

First, we note that the path $\Gamma_{X}$ at $t_i$ and $t_f$ is smoothly deformed to topologically distinct path $\Gamma_{X}$ and $\Gamma'_{X}$ at some intermediate moment $t_m$. Fig. \ref{fig_SMVII}(b) displays the three relevant paths, they satisfy:
\begin{eqnarray}\label{bxytm}
b_{\Gamma'_{X}}(t_m)=b_{\Gamma_{Y}}^{-1}(t_m)b_{\Gamma_{X}}(t_m)b_{\Gamma_{Y}}(t_m).
\end{eqnarray}
According to dynamical NACR, we have
\begin{eqnarray}\label{dym}
b_{\Gamma_{X/Y}}(t_m)=b_{dyn}^{-1}(t_i\rightarrow t_m)b_{\Gamma_{X/Y}}(t_i)b_{dyn}(t_i\rightarrow t_m).
\end{eqnarray}
Inserting Eq. (\ref{dym}) into Eq. (\ref{bxytm}), we obtain
\begin{eqnarray}\label{bxytm2}
b_{\Gamma'_{X}}(t_m)=b_{dyn}^{-1}(t_i\rightarrow t_m)b_{\Gamma_{Y}}^{-1}(t_i)b_{\Gamma_{X}}(t_i)b_{\Gamma_{Y}}(t_i)b_{dyn}(t_i\rightarrow t_m).
\end{eqnarray}
The path $\Gamma_{X}$ at the final stage $t_f$ is topologically equivalent to the path $\Gamma'_{X}$, according to dynamical NACR,
\begin{eqnarray}\label{bgmf}
\begin{split}
b_{\Gamma_{X}}(t_f)&=b_{dyn}^{-1}(t_m\rightarrow t_f)b_{\Gamma'_{X}}(t_m)b_{dyn}(t_m\rightarrow t_f)\\
&=b_{dyn}^{-1}(t_i\rightarrow t_f)  b_{\Gamma_{Y}}^{-1}(t_i)b_{\Gamma_{X}}(t_i)b_{\Gamma_{Y}}(t_i)  b_{dyn}(t_i\rightarrow t_f),
\end{split}
\end{eqnarray}
where we have used $b_{dyn}(t_i\rightarrow t_f)=b_{dyn}(t_i\rightarrow t_m) b_{dyn}(t_m\rightarrow t_f)$. By appropriately choosing base point $\bm{P}$ such that $b_{dyn}(t_i\rightarrow t_f)=1$, we arrive at the conclusion:
\begin{eqnarray}
b_{\Gamma_{X}}(t_f)= b_{\Gamma_{Y}}^{-1}(t_i)b_{\Gamma_{X}}(t_i)b_{\Gamma_{Y}}(t_i).
\end{eqnarray}
In particular, if $b_X(t_i)$ and $b_Y(t_i)$ commute, $b_X(t_i)b_Y(t_i)=b_Y(t_i)b_X(t_i)$, we have $b_{\Gamma_{X}}(t_f)=b_{\Gamma_{X}}(t_i)$. That is, if the targeted EP locus over-crosses another commutative EP locus, the braiding of the targeted EP does not chang; otherwise it will shift to its conjugate partner. For the color change shown in Fig. 2(c) of the main text, when the red EP locus with braiding $\tau_1^{-1}$ under-crosses the non-commutative orange EP locus with braiding $\tau_2$, the braiding of the red EP changes from $\tau_1^{-1}$ to $\tau_2^{-1}\tau_1^{-1}\tau_2$. Similarly, when the green EP locus with braiding $\tau_2^{-1}\tau_1^{-1}\tau_2$ under-crosses the non-commutative blue EP locus with braiding $\tau_1$, the braiding changes from $\tau_2^{-1}\tau_1^{-1}\tau_2$ to $\tau_1^{-1}\tau_2^{-1}\tau_1^{-1}\tau_2\tau_1=\tau_2^{-1}$. For the no-go transition shown in Fig. 3(a) of the main text, when the red EL with braiding $b_1$ under-crosses the non-commutative blue EL with braiding $b_2$, the braiding of the red EL changes from $b_1$ to $b'_1=b_2^{-1}b_1b_2$.

\subsection{(IX) Model realizations of the EL configurations of Fig. 3 in acoustic cavities}
\begin{figure*}[!h]
\includegraphics[width=.7\textwidth]{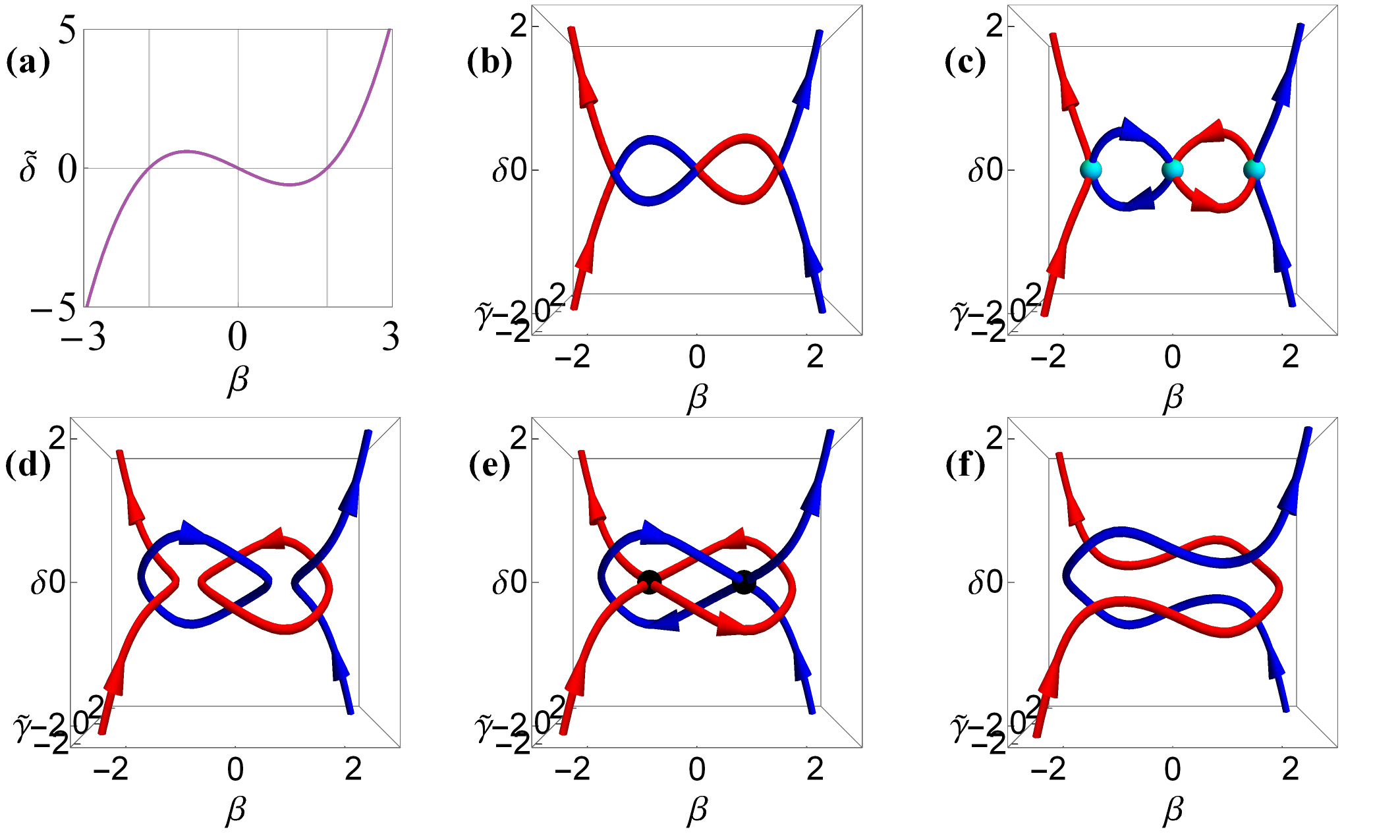}
\caption{Detailed model realizations of the EL configurations of Fig. 3 in the main text. The Hamiltonian is given by Eq. (\ref{H3D_SM}). (a) Parameter $\tilde{\delta}(\beta)=0.3((\beta-1)^3+3(\beta-1)^2-2)$ as a function of $\beta$. (b-f) The EL configurations in the 3D $(\beta,\tilde{\gamma},\delta)$ space with $\gamma(t)=t, \tilde{\delta}(\beta)=0.3((\beta-1)^3+3(\beta-1)^2-2)$ for $t=1.5, t=1, t=0.6, t=0.574, t=0.4$, respectively. The blue curves represent ELs from the degeneracy of $E_1$ and $E_2$, while the red curves represent ELs from the degeneracy of $E_2$ and $E_3$. The cyan dots represent the third-order EPs.}
\label{fig_SMVIII_1}
\end{figure*}

In the main text, we have shown several examples of transitions between different EL configurations in Fig. 3 allowed by the NACR. Here we illustrate a concrete three-band model to realize these ELs. The Hamiltonian reads
\begin{equation}\label{H3D_SM}
H=\left(
\begin{array}{ccc}
\sqrt{2}i[\gamma(t)-i\delta] & -\kappa & 0 \\
-\kappa & i[-\tilde{\gamma}+i\tilde{\delta}(\beta)] & -\kappa \\
0 & -\kappa & -\sqrt{2}i[\gamma(t)-i\delta] \\
\end{array}
\right).
\end{equation}
The model can be easily realized using the same coupled acoustic cavities in Fig. 2b of the main text. Still, we set the coupling strength between neighboring cavities as $\kappa=1$. $\gamma(t)$, $-\tilde{\gamma}$, $-\gamma(t)$ $\in\mathbb{R}$ are the gain or loss in the respective cavities, $\delta$, $\tilde{\delta}(\beta)$, $-\delta$ $\in\mathbb{R}$ are the detunings. We set $\tilde{\delta}(\beta)=0.3((\beta-1)^3+3(\beta-1)^2-2)$ ($\beta\in\mathbb{R}$ ) as depicted in Fig. \ref{fig_SMVIII_1}(a). Here, we slowly vary the parameters $\gamma(t)=t$ and examine the evolution of ELs in the 3D $(\beta,\tilde{\gamma},\delta)$ space.

\begin{figure*}[btp]
\includegraphics[width=.7\textwidth]{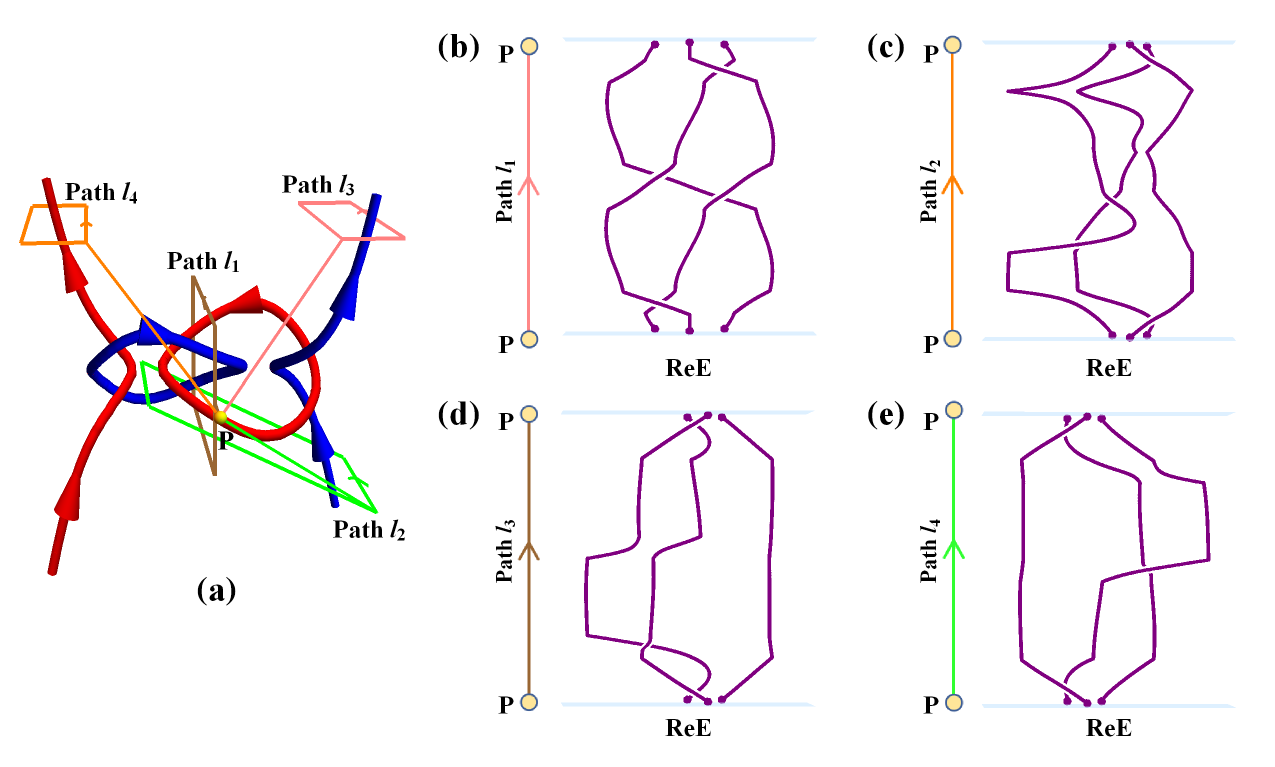}
\caption{Braid invariants of the staggered-ring ELs for four representative paths $l_1$, $l_2$, $l_3$ and $l_4$ based at $\bm{P}$. (a) The EL configurations of Hamilonian (\ref{H3D_SM}) in the 3D $(\beta,\tilde{\gamma},\delta)$ space with $\gamma(t)=t, \tilde{\delta}(\beta)=0.3((\beta-1)^3+3(\beta-1)^2-2)$ for $t=0.6$. (b-e) The eigenvalue braidings along the four paths $l_1$, $l_2$, $l_3$ and $l_4$ in (a), respectively. The base point $\bm{P}$ is pinned at $(\beta,\tilde{\gamma},\delta)=(0,-2.5,0)$ as given by yellow dot in (a).}
\label{fig_SMVIII_2}
\end{figure*}

The characteristic polynomial of the Hamiltonian (\ref{H3D_SM}) is obtained as
\begin{equation}
    f(\delta,\tilde{\gamma})=\mathrm{det}[E-H]=E^3+g_1E^2+g_2E+g_3,
\end{equation}
with
\begin{equation}
\begin{array}{lll}
   g_1=\tilde{\delta}+i\tilde{\gamma},\\
   g_2=2(\gamma^2-\delta^2-1-2i\gamma\delta),\\
   g_3=2(2\gamma\tilde{\gamma}\delta+\gamma^2\tilde{\delta}-\delta^2\tilde{\delta}-i\tilde{\gamma}\delta^2-2i\gamma\delta\tilde{\delta}+i\gamma^2\tilde{\gamma}).
\end{array}
\end{equation}
Using the same Sylvester matrix method, we can determine the EL configurations of Hamiltonian (\ref{H3D_SM}) in the 3D $(\beta,\textcolor{red}{\tilde{\gamma}},\delta)$ space. As demonstrated in Figs. \ref{fig_SMVIII_1}(b)-(f), we produce the same ELs in Fig. 3c of the main text, corresponding to $t=1.5, t=1, t=0.6, t=0.574, t=0.4$, respectively. Instead of using different colors to represent different components of ELs in the main text, here we use the blue color to represent ELs from the degeneracy of $E_1$ and $E_2$, and the red color to represent ELs from the degeneracy of $E_2$ and $E_3$. Notably, the color of the ELs changes suddenly for each continuous component in Fig. \ref{fig_SMVIII_1}(b), which further confirms the failure of a globally consistent numbering of eigenvalues. In addition, when the condition $g_1=g_2=g_3=0$ is satisfied, all three eigenvalues are degenerate at $E=0$. This condition is satisfied when $\tilde{\delta}=\tilde{\gamma}=\delta=0,\gamma=1$, and $\tilde{\delta}=0$ further requires $\beta=-\sqrt{3}, 0, \sqrt{3}$, at which all three eigenvectors coalesce into $[-1,-i\sqrt2,1]^T$. Therefore, $(\beta,\tilde{\gamma},\delta)=(-\sqrt{3},0,0)$, $(\beta,\tilde{\gamma},\delta)=(0,0,0)$, and $(\beta,\tilde{\gamma},\delta)=(\sqrt{3},0,0)$ are third-order $\mathrm{EPs}$ (cyan dots) when $t=1$, as illustrated in Fig. \ref{fig_SMVIII_1}(c).

Now let us turn to the transitions between different EL configurations from the perspective of the dynamical NACR. We choose four representative paths $l_1,l_2,l_3,l_4$, which do not touch any ELs during the whole evolution. One can check that the braid invariants along these four paths remain unchanged, in agreement with the dynamical NACR. In Fig. \ref{fig_SMVIII_2}, we plot the eigenvalue braidings along the four paths when $t=0.6$. The braid invariants along the four representative paths $l_1,l_2,l_3,l_4$ are obtained as $b_{l_1}=\tau_1^{-1}\tau_2\tau_1\tau_2^{-1}$, $b_{l_2}=1$, $b_{l_3}=\tau_1$, $b_{l_4}=\tau_1^{-1}\tau_2\tau_1$. We notice that the braid invariant along the path $l_2$ is trivial $b_{l_2}=1$; thus, the second and fourth components (counted from left to right) are marked in the same color in Fig. 3 of the main text. In addition, the braid invariant along the path $l_1$ is non-trivial $b_{l_1}=\tau_1^{-1}\tau_2\tau_1\tau_2^{-1}$, which forbids the untying of the two central components in Fig. \ref{fig_SMVIII_2}(d).

\subsection{(X) Example of the EL emergence allowed by the NACR}
\begin{figure*}[h]
\includegraphics[width=.7\textwidth]{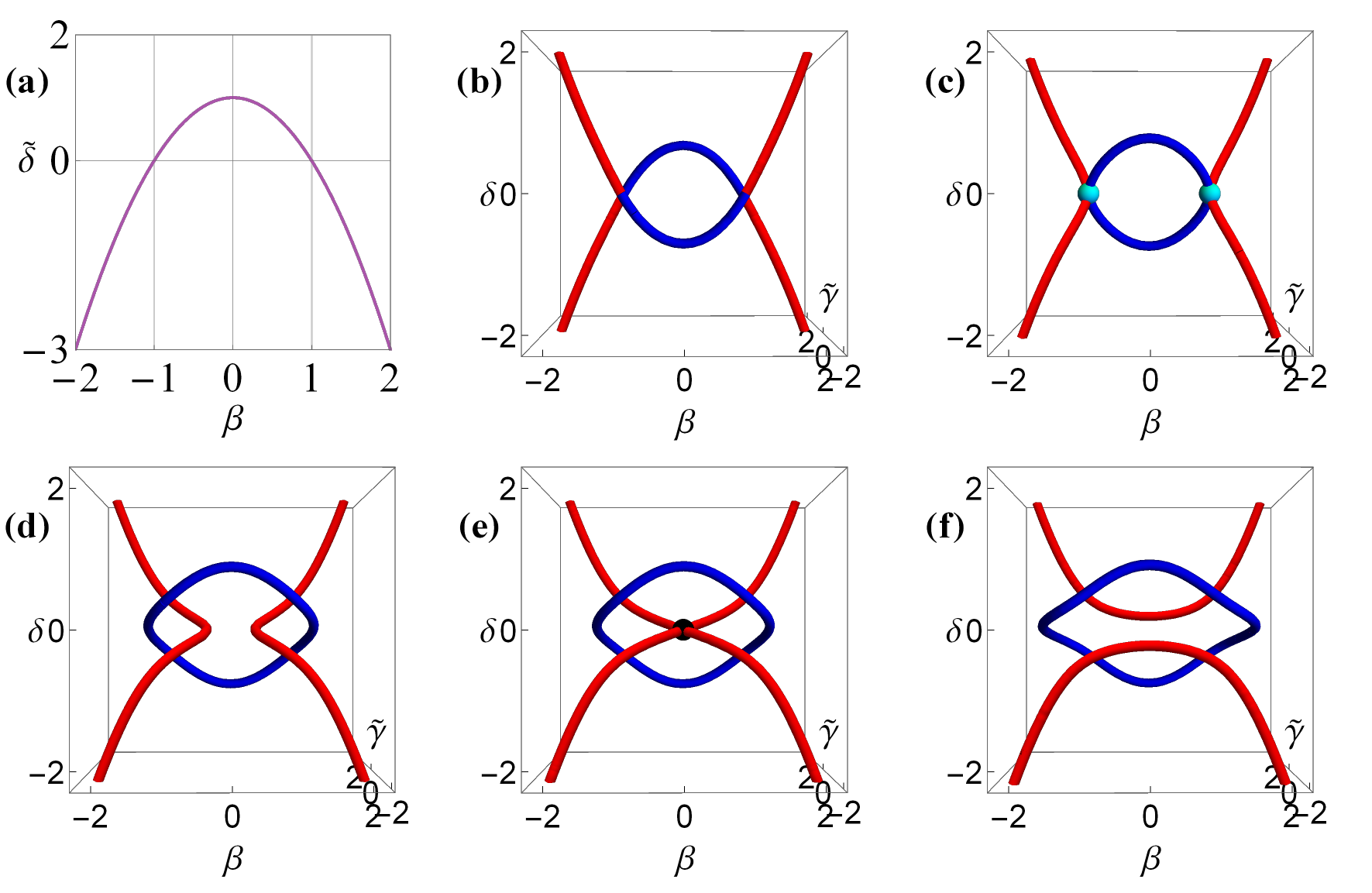}
\caption{An example of the emergence of an exceptional ring permitted by the NACR. (a) Parameter $\tilde{\delta}(\beta)=-\beta^2+1$ as a function of $\beta$. (b-f) The EL configurations of the Hamiltonian (\ref{H3D_SM}) in the 3D $(\beta,\tilde{\gamma},\delta)$ space with $\gamma(t)=t, \tilde{\delta}(\beta)=-\beta^2+1$ for $t=1.5, t=1, t=0.5, t=0.458, t=0.3$, respectively. The blue curves represent ELs from the degeneracy of $E_1$ and $E_2$, while the red curves represent ELs from the degeneracy of $E_2$ and $E_3$. The cyan dots represent third-order EPs.}
\label{fig_SMIX}
\end{figure*}

In the main text, we have demonstrated how to analyze the admissible EL configurations through the NACR. Here we provide more examples. We still take the Hamiltonian Eq. (\ref{H3D_SM}) as example and consider the case of $\tilde{\delta}(\beta)=-\beta^2+1$, as depicted in Fig. \ref{fig_SMIX}(a). We plot the EL configurations in the 3D $(\beta,\tilde{\gamma},\delta)$ space with $t=1.5, t=1, t=0.5, t=0.458, t=0.3$ in Figs. \ref{fig_SMIX}(b)-(f), respectively. When the condition $g_1=g_2=g_3=0$ is satisfied, all three eigenvalues are degenerate at $E=0$. This condition is satisfied when $\tilde{\delta}=\tilde{\gamma}=\delta=0,\gamma=1$, and $\tilde{\delta}=0$ further requires $\beta=-1, 1$, at which all three eigenvectors coalesce into $[-1,-i\sqrt2,1]^T$. Therefore, $(\beta,\tilde{\gamma},\delta)=(-1,0,0)$ and $(\beta,\tilde{\gamma},\delta)=(1,0,0)$ are third-order $\mathrm{EPs}$ (cyan dots) when $t=1$, as shown in Fig. \ref{fig_SMIX}(c).

Starting from the two initially untangled ELs [Fig. \ref{fig_SMIX}(b)] when $t=1.5$, we observe that the two ELs gradually approach each other as $t$ decreases and touch at two third-order EPs [Fig. \ref{fig_SMIX}(c)] when $t=1$. As $t$ continues to decrease, there emerges an additional exceptional ring through the recombination of the ELs [Fig. \ref{fig_SMIX}(d)]. When $t=0.3$, the ELs become untangled again [Fig. \ref{fig_SMIX}(f)]. Comparing (b) with (f), the whole process can be regarded as the emergence of the exceptional ring. One can similarly check the validity of the NACR by focusing on the eigenvalue braidings along several representative paths in the parameter space.

\subsection{(XI) Comparison between different topological invariants}
In this subsection, we compare different topological invariants in characterizing non-Hermitian systems. The braid group $B_N$ is the fundamental group of the classifying space of $N$-band non-Hermitian Hamiltonians. It is rigorously obtained from the homotopy theory and fully captures the non-Abelian topology of non-Hermitian systems. The permutation group $S_N$ is a finite subgroup of $B_N$ and generated by adjacent transpositions as below:
\begin{eqnarray}
S_N=\langle s_1,...,s_{N-1}:~s_is_{i+1}s_i=s_{i+1}s_is_{i+1},s_is_j=s_js_i ~\mathrm{for}~ |i-j|>1,~\mathrm{and}~s_i^2=1\rangle.
\end{eqnarray}
The permutation group lacks some key information about the exceptional physics. In braid group, the two elements $\tau_i$ and $\tau_i^{-1}$ represent over- and under-crossings of the $i$-th and $(i+1)$-th energy levels. They correspond to the same transposition in the permutation group $(i,i+1)$. That is, the permutation group only cares about the transposition of two levels, regardless of whether it is over- or under-crossing. Another widely-used topological invariant in non-Hermitian systems is the discriminant number $D_{\Gamma}$. As an integer invariant constructed from the characteristic polynomial, it only captures the net spectral windings and cannot scrutinize the non-Abelian nature of exceptional degeneracies. For the braid invariant $\tau_{i_1}^{b_1}\tau_{i_2}^{b_2}...\tau_{i_q}^{b_q}$ associated with some closed path $\Gamma$, the corresponding discriminant number is $D_{\Gamma}=b_1+b_2+...+b_q$. For instance, the discriminant number cannot distinguish the two topological distinct paths with braid invariants $\tau_1^2\tau_3^{-1}$ and $\tau_1$.
\begin{figure*}[h]
\includegraphics[width=.5\textwidth]{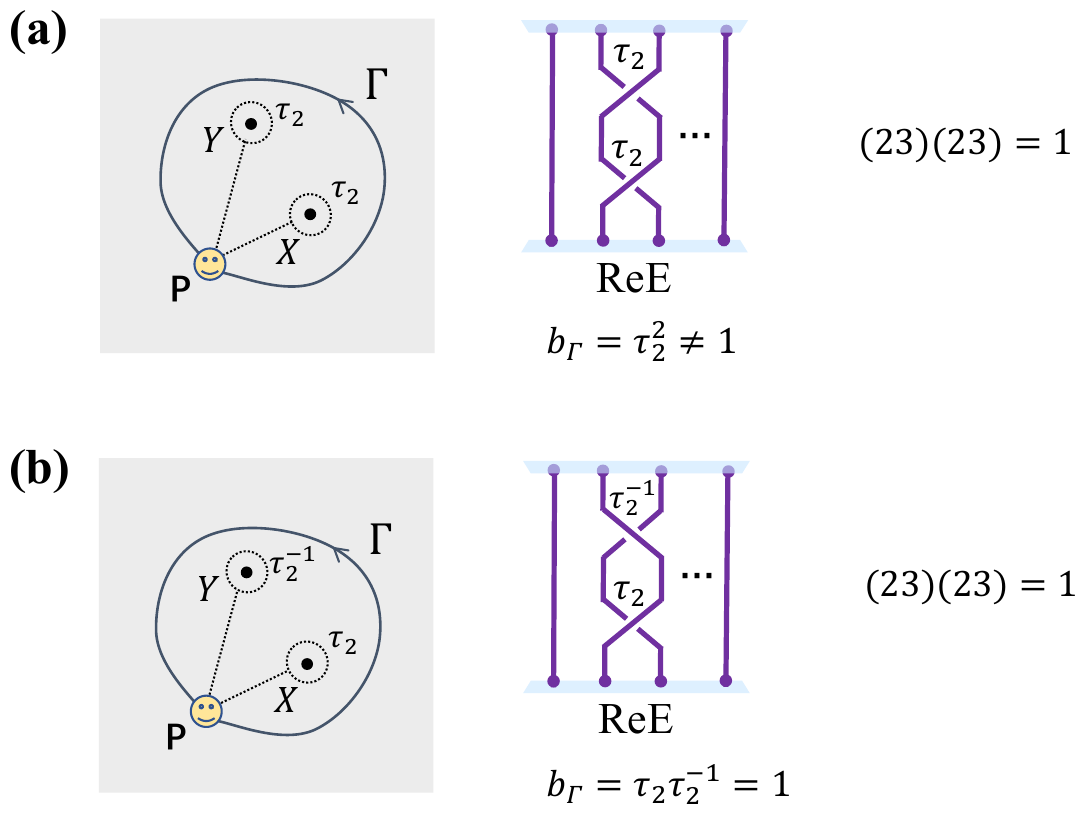}
\caption{ Comparison between braid invariant and permutation invariant. Left panel: a closed path $\Gamma$ enclosing two EPs (labeled as $X$, $Y$) between the second and third energy level. The local braid invariants of $X$ and $Y$ are $b_X=b_Y=\tau_2$ in (a); $b_X=\tau_2$, $b_Y=\tau_2^{-1}$ in (b). Middle panel: the braid diagram of energy levels along $\Gamma$. Right panel: state permutation along $\Gamma$.}
\label{fig_SMNew2}
\end{figure*}

The collective behaviors of multiple EPs/ELs (like the emergence, annihilation, coalescence, and braiding of EPs/ELs as discussed in the main text) and the interplay between different EPs/ELs are fully governed by the braid-group invariant. It records all the necessary information of non-Abelian topology and avoids the ambiguity or oversimplification of the permutation group or discriminant number. This is best illustrated by the example depicted in Fig. \ref{fig_SMNew2}. Let us consider two EPs (denoted as $X$ and $Y$) with local braid invariants $b_X=\tau_2$, $b_Y=\tau_2$ in Fig. \ref{fig_SMNew2}(a) and $b_X=\tau_2$, $b_Y=\tau_2^{-1}$ in Fig. \ref{fig_SMNew2}(b). In both cases, $X$ and $Y$ correspond to the exchange of the second and third energy levels. They have the same permutation invariant $(23)$. To predict whether the two EPs will annihilate each other when brought together, we examine the braid invariant $b_{\Gamma}$ along the outer closed path $\Gamma$, which encloses both EPs. It turns out the braid invariants are respectively $b_{\Gamma}=\tau_2^2\neq 1$ and $b_{\Gamma}=\tau_2\tau_2^{-1}=1$ in Figs. \ref{fig_SMNew2}(a)(b). That is, the two EPs cannot/can annihilate each other in Fig. \ref{fig_SMNew2}(a)/(b). However, the permutation invariant along the closed path $\Gamma$ is trivial: $(23)(23)=1$ for both cases. Hence the permutation group is unable to distinguish the two cases.

\end{document}